\documentclass[%
 reprint,
 superscriptaddress,
 amsmath,amssymb,
 aps,
]{revtex4-2}

\usepackage{graphicx}
\usepackage{dcolumn}
\usepackage{bm}
\usepackage{hyperref}
\usepackage{physics}
\usepackage{xcolor}
\usepackage{mathtools}
\usepackage{amssymb}
\usepackage{color,soul}
\usepackage{float}

\usepackage{mathtools}

\DeclarePairedDelimiter\floor{\lfloor}{\rfloor}

\usepackage{accents}

\newcommand\varpm{\mathbin{\vcenter{\hbox{%
				\oalign{\hfil$\scriptstyle+$\hfil\cr
					\noalign{\kern-.3ex}
					$\scriptscriptstyle({-})$\cr}%
}}}}
\newcommand\varmp{\mathbin{\vcenter{\hbox{%
				\oalign{$\scriptstyle({+})$\cr
					\noalign{\kern-.3ex}
					\hfil$\scriptscriptstyle-$\hfil\cr}%
}}}}

\begin{document}


\title{Topological bound states in a lattice of rings with nearest-neighbour interactions}

\author{Yunjia Zhai}
\email{yunjia.zhai@autonoma.cat}
\affiliation{Departament de F\'{i}sica, Universitat Aut\`{o}noma de Barcelona, 08193 Bellaterra, Spain}

\author{Ayaka Usui}
\email{ayaka.usui@uab.cat}
\affiliation{Departament de F\'{i}sica, Universitat Aut\`{o}noma de Barcelona, 08193 Bellaterra, Spain}

\author{Anselmo M. Marques}
\affiliation{Department of Physics and i3N, University of Aveiro, 3810-193 Aveiro, Portugal}

\author{Ricardo G. Dias}
\affiliation{Department of Physics and i3N, University of Aveiro, 3810-193 Aveiro, Portugal}

\author{Ver\`{o}nica Ahufinger}
\email{veronica.ahufinger@uab.cat}
\affiliation{Departament de F\'{i}sica, Universitat Aut\`{o}noma de Barcelona, 08193 Bellaterra, Spain}
\affiliation{ICFO - Institut de Ciències Fotòniques, The Barcelona Institute of Science and Technology,
08860 Castelldefels (Barcelona), Spain}


\begin{abstract}
We study interaction-induced bound states in a system of ultracold bosons loaded into the states with orbital angular momentum in a one-dimensional staggered lattice of rings. We consider the hard-core limit and strong nearest-neighbour interactions such that two particles in next neighbouring sites are bound. Focusing on the manifold of such bound states, we have derived the corresponding effective model for doublons. With orbital angular momentum $l=1$, the original physical system is described as a Creutz ladder by using the circulations as a synthetic dimension, and the effective model obtained consists of two Su-Schrieffer-Heeger (SSH) chains and two Bose-Hubbard chains. Therefore, the system can exhibit topologically protected edge states. In a structure that alternates $l=1$ and $l=0$ states, the original system can be mapped to a diamond chain. In this case, the effective doublon model corresponds to a Creutz ladder with extra vertical hoppings between legs and can be mapped to two SSH chains if all the couplings in the original system are equal. Tuning spatially the amplitude of the couplings destroys the inversion symmetry of these SSH chains, but enables the appearance of multiple flat bands.
\end{abstract}

\maketitle


\section{Introduction}

Non-trivial topological systems are characterized by exhibiting robust states protected against local perturbations, as long as the relevant symmetries that define their topological nature remain preserved~\cite{Hasan2010Colloquium}. The interest of this type of systems has been extended from condensed matter to other fields. In particular, topologically protected states have been reported in platforms such as ultracold atoms~\cite{Goldman2016Topological,Cooper2019Topological}, photonic systems~\cite{Lu2014Topological,Ozawa2019Topological,Leykam2020Topological,Price2022,Mehrabad2023}, electrical circuits~\cite{Lee2018,Wang2020Circuit} and acoustic lattice~\cite{Yang2015,Zhang2018,Xue2022}. 

The topological characterization of single-particle systems is well established~\cite{Qi2011,Chiu2016}, but its extension to interacting systems still presents open questions. Thus, the study of topology in interacting few-body and many-body systems has attracted a lot of attention in different contexts and has revealed rich physics~\cite{Rachel2018,Fidkowski2010Effects,Xie2012Symmetry,Lin2023Topological,deLsleuc2019,Martinez2023,Xie2020,Maciejko2013,Budich2013}. Of particular interest are the bound states of two particles, known as doublons~\cite{Mattis1986}, which have been experimentally observed in ultra-cold atom systems in optical lattices~\cite{Flling2007, Winkler2006Winkler,Preiss2015,Tai2017}, in photonic systems ~\cite{Lahini2012, Mukherjee2016}, in topolectrical circuits~\cite{Olekhno2020} and in superconducting circuits~\cite{Martinez2023}. 

The topological properties of doublons have been extensively investigated in various lattice models such as the SSH model~\cite{Liberto2016Two,Bello2016,Liberto2017Two,Marques2017,Gorlach2017,Marques2018,Gorlach2018,Azcona2021}, the Creutz model~\cite{Zurita2020,Kuno2020,pelegri2024few}, the Haldane model~\cite{Salerno2018}, the extended Hubbard model~\cite{Gorlach2017b,Bello2017,Lyubarov2019,Lin2020,Stepanenko2020,Ling2025,Su2025}, the diamond lattice~\cite{Pelegr2020Interaction}, the Rice-Mele model~\cite{Ke2017}, the XXZ model~\cite{Qin2017}, the Hatano-Nelson model~\cite{Brighi2024}, a ladder lattice~\cite{Zheng2023Two} or lattices with spin-orbit coupling~\cite{Guo2011}. There are several approaches to explore the topological features of doublons. An effective strategy involves treating doublons as quasiparticles, which allows for the application of topological phase classification techniques developed for single-particle systems~\cite{Guo2011,Marques2017,Marques2018, Gorlach2017,Zurita2020,Salerno2020Interaction}. Alternatively, one-dimensional (1D) interacting systems can be mapped onto non-interacting two-dimensional (2D) models~\cite{Liberto2016Two,Gorlach2017,Gorlach2018}. Moreover, it is possible to construct effective lattice geometries for the doublon center-of-mass~\cite{Salerno2020Interaction,Marques2018}. 

In this work, we investigate a system of doublons induced by nearest-neighbour (NN) interactions in a 1D staggered lattice of ring potentials with alternating distances.
Each ring potential has eigenstates with orbital angular momentum (OAM) $l$ with winding number $\pm l$ and the system exhibits complex couplings that can be engineered by modifying the geometry of the lattice~\cite{Polo2016Geometrically,Pelegri2019Topological,Pelegri2019Topological2,Pelegr2020Interaction, Pelegri2019Second, Pelegri2019Quantum,Nicolau2023Many, Nicolau2024Ultracold}. With $l=0$ at all sites, our physical system reduces to the SSH model, which is the simplest topological insulator~\cite{Su1979Solitons}. With $l=1$, each site has two states corresponding to the two circulations, and the 1D lattice can be mapped to a Creutz ladder~\cite{RevModPhys.73.119} using the circulations as a synthetic dimension~\cite{nicolau2023bosonic,Nicolau2023Many}. 
We have derived the effective model for doublons in the case of $l=1$ with infinite repulsive on-site interactions and in the limit of large NN interactions. We have found that it consists of four decoupled chains, including two SSH chains and two Bose-Hubbard chains. Therefore, topological edge states appear in some parameter regimes. 
In addition, bosons can also be loaded with different OAM $l=0$ and $l=1$ at alternating sites. Then, the lattice spanned in both real and synthetic dimensions has a diamond structure. 
The effective model of doublons in this system corresponds to a Creutz ladder with extra inter-leg hoppings, which can also be converted to two SSH chains if all the couplings in the original system are equal.
Unlike the previously studied single-particle Creutz ladder~\cite{PhysRevX.7.031057}, this extended Creutz model for doublon features staggered inter-leg couplings, generating four flat bands and hosting edge states embedded in the flat band. The complex diagonal and horizontal couplings create four distinct $\pi$ flux loops. Our work provides a method for the realization of an extended Creutz ladder in a lattice of rings, giving access to the topological states and multiple flat-band structure investigated here.
 If these couplings vary spatially, the effective model deviates from the SSH chains, and the band structure may exhibit multiple flat bands.

This paper is organised as follows.
In Sec.~\ref{sec:physicalsystem}, we introduce the ring lattice system and the Hamiltonian for the cases of $l=1$ in all the sites and of $l=0$ and $l=1$ in alternating sites. In Sec.~\ref{sec:bound}, we impose strong NN interactions as well as infinite repulsive on-site interactions, and we derive the effective Hamiltonians for doublons in the two considered cases in Sec.~\ref{sec:doublon_l1} and Sec.~\ref{sec:doublon_l01}, respectively, and characterise them topologically. 
We present our conclusions in Sec.~\ref{sec:conclusions}.

\section{Physical system} \label{sec:physicalsystem}

\begin{figure}[t]
    \includegraphics[width=1.0\linewidth]{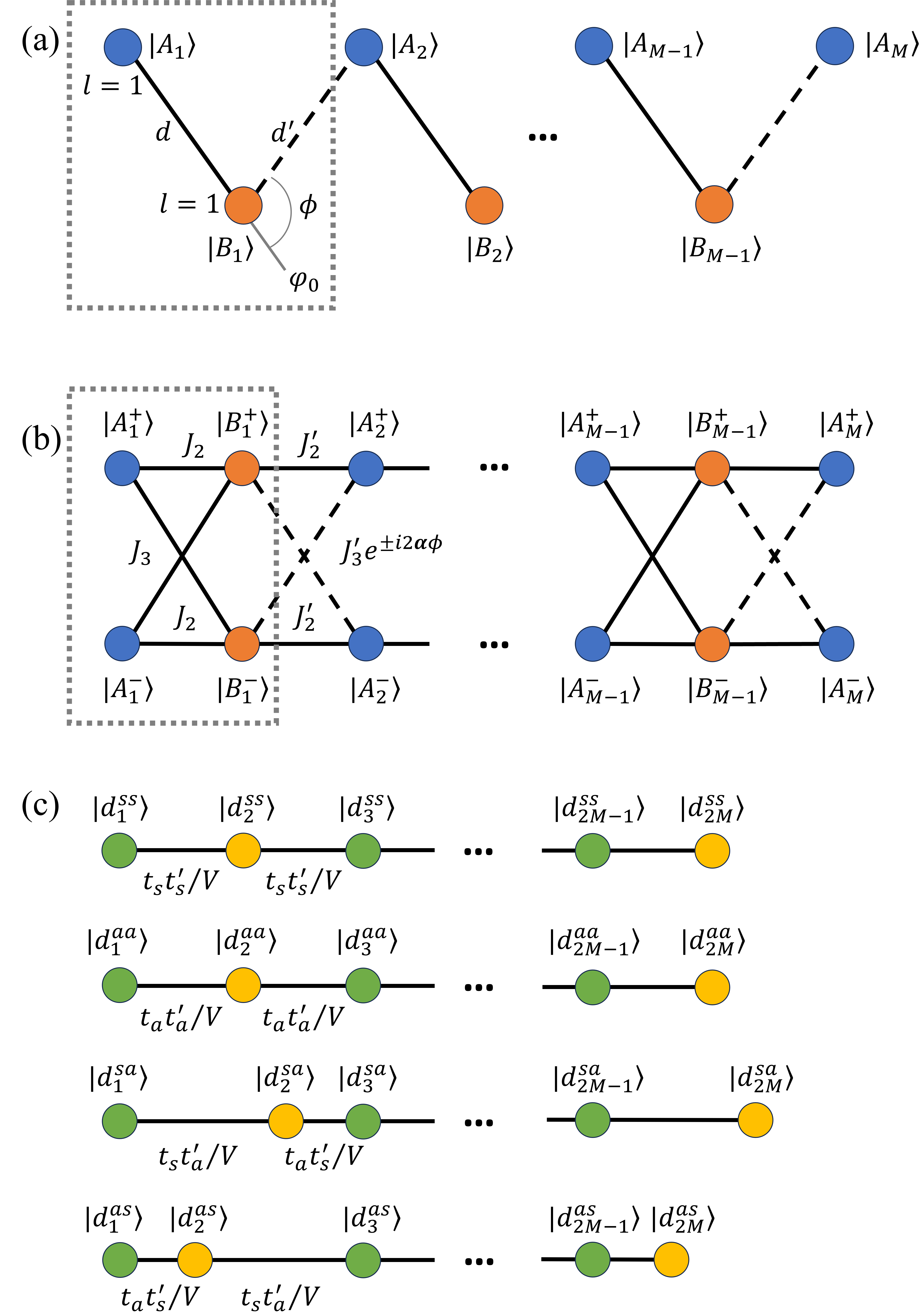}
    \caption{(a) Sketch of the 1D staggered chain of rings loaded with bosons with OAM $l=1$. The chain ends with an $A$ site, and thus the number of unit cells is $N_c=M+1/2$ with $M\in\mathbb{N}$. (b) Sketch of the Creutz ladder formed by the real dimension and the synthetic dimension spanned by two circulations $\pm$ in each site $A_j$ and $B_j$. (c) Sketch of the four decoupled chains that correspond to the effective doublon model in Eq.~\eqref{eq:H_eff_l1}. The number of sites here is $2N_c-1=2M$. The solid lines indicate real tunneling amplitudes, and the dashed lines indicate complex tunneling amplitudes between states with OAM modes with different circulations. The unit cells in (a) and (b) are marked by grey dotted rectangles.
    The distance between sites is reflected in the relative magnitude of the hopping terms.
    }
    \label{fig:l1model}
\end{figure}

We consider a 1D staggered lattice of ring potentials loaded with bosons in OAM states~\cite{nicolau2023bosonic,Nicolau2023Many,Nicolau2024Ultracold}.
Below, we introduce two different configurations and give the general formulation of their Hamiltonians. 

\subsection{Case of $l=1$}

In this subsection, we consider the staggered lattice of rings shown in Fig.~\ref{fig:l1model}(a). The pair of sites $A_j$ and $B_j$ forms the unit cell $j$. The intracell and intercell distances are denoted by $d$ and $d'$, respectively. Note that the angle $\phi$ plays an important role to induce complex couplings~\cite{Polo2016Geometrically,Pelegri2019Topological,Pelegr2020Interaction,Pelegri2019Topological2,Pelegri2019Quantum,Pelegri2019Second}, as detailed below. The eigenstates of each isolated ring have a well defined OAM $l=1$ with winding numbers $\gamma=\pm 1$ and circulations $\alpha=\pm$. 
The total Hamiltonian is expressed as
\begin{align} \label{eq:H_total}
    \hat{H}_{l=1}
    &=
    \hat{H}_{l=1}^0
    +
    \hat{H}_{l=1}^{U}
    +
    \hat{H}_{l=1}^{V}
    ,
\end{align}
where $\hat{H}_{l=1}^0$ is the single-particle Hamiltonian, $\hat{H}_{l=1}^{U}$ is the on-site interaction Hamiltonian, and $\hat{H}_{l=1}^{V}$ is the NN interaction Hamiltonian. The single-particle Hamiltonian $\hat{H}_{l=1}^0$ reads
\begin{align} \label{eq:H_l1_sp}
    \hat{H}_{l=1}^0
    &=
    \sum_{\alpha=\pm}
    \bigg(
    J_2 
    \sum_{j=1}^{N_c}
    \hat{a}_j^{\alpha\dagger} 
    \hat{b}_j^{\alpha}
    +
    J'_2 
    \sum_{j=1}^{N_c-1}
    \hat{b}_j^{\alpha\dagger} 
    \hat{a}_{j+1}^{\alpha}
    \nonumber\\
    &\quad
    +
    J_3
    \sum_{j=1}^{N_c}
    \hat{a}_j^{\alpha\dagger} 
    \hat{b}_j^{-\alpha}
    +
    J'_3 e^{-i2\alpha\phi}
    \sum_{j=1}^{N_c-1}
    \hat{b}_j^{\alpha\dagger} 
    \hat{a}_{j+1}^{-\alpha}
    \bigg)
    \nonumber\\
    &\quad
    +\text{H.c.}
    ,
\end{align}
where $\hat{a}_j^{\alpha }$ $(\hat{a}_j^{\alpha \dagger})$ and $\hat{b}_j^{\alpha }$ $(\hat{b}_j^{\alpha \dagger})$ are the annihilation (creation) operators of state $\ket{m_j^{\alpha}}$, where $j$ is the unit cell index, $m=A,B$ labels the site, and $\alpha$ is the circulation. The number of unit cells is $N_c$. Note that the cell number $N_c$ can be an integer but also a half integer, and the latter means that the chain ends with an $A$ site. The coefficient $J_2$ ($J_2'$) is the intracell (intercell) hopping strength between OAM modes with the same circulation, and the coefficient $J_3$ ($J_3'$) is the intracell (intercell) hopping strength between OAM modes with different circulations. The complex tunneling phases come from the azimuthal phase of the OAM modes on each ring when calculating the overlap integral between the wavefunctions associated with local eigenstates of neighbouring rings.
Particularly, the intercell hopping between $\pm$ modes contains a complex factor $e^{-i2\alpha\phi}$~\cite{Polo2016Geometrically,Nicolau2023Many}, where the staggering angle $\phi$ can be tuned. 
This system can be depicted as a Creutz ladder by using the circulations as a synthetic dimension, i.e., representing the two circulations as separate sites [see Fig.~\ref{fig:l1model}(b)].
Besides the four hopping terms that appear in Eq.~\eqref{eq:H_l1_sp}, there are possibilities where other hopping terms also come up. For instance, the hopping between $A_j$ and $A_{j+1}$ (and between $B_j$ and $B_{j+1}$) cannot be neglected if the distance between these sites, which depends on the angle $\phi$, is comparable to $d,d'$ (see Fig.~\ref{fig:l1model}(a)). We prevent this by restricting the angle to the $\phi\in[0,2\pi/3)$ region, where A–A and B–B couplings are effectively suppressed
~\cite{Polo2016Geometrically}. Moreover, a coupling between opposite winding number OAM modes within a single ring (i.e. hopping between $\ket{A_j^+}$ and $\ket{A_j^-}$ and between $\ket{B_j^+}$ and $\ket{B_j^-}$) is also present in the system due to the breaking of the cylindrical symmetry. Nevertheless, it has been shown~\cite{Pelegri2019Topological} that the amplitude of this coupling is one order of magnitude smaller than the coupling strengths in Eq.~\eqref{eq:H_l1_sp}, $J_2$ and $J_3$, and thus we neglect it.

We consider the symmetric and the antisymmetric superpositions of the two circulations in each site of the lattice, 
\begin{subequations}\label{eq:basis_sa_0}
\begin{align}
\begin{aligned}
    \ket{A_j^{s}}
    &\equiv \frac{1}{\sqrt{2}}\left(\ket{A_j^+} + \ket{A_j^-}\right), \\
    \ket{A_j^{a}}
    &\equiv \frac{1}{\sqrt{2}}\left(\ket{A_j^+} - \ket{A_j^-}\right),
\end{aligned}
\end{align}
\begin{align}
\begin{aligned}
    \ket{B_j^{s}}
    &\equiv \frac{1}{\sqrt{2}}\left(\ket{B_j^+} + \ket{B_j^-}\right), \\
    \ket{B_j^{a}}
    &\equiv \frac{1}{\sqrt{2}}\left(\ket{B_j^+} - \ket{B_j^-}\right).
\end{aligned}
\end{align}
\end{subequations}
and then define the corresponding creation and annihilation operators as 
\begin{subequations}
\label{eq:basis_sa}
\begin{align}
\hat a_j^{\,s}
&\equiv
\frac{1}{\sqrt{2}}
\left(
\hat a_j^{+} + \hat a_j^{-}
\right),
&
\hat a_j^{\,a}
&\equiv
\frac{1}{\sqrt{2}}
\left(
\hat a_j^{+} - \hat a_j^{-}
\right),
\\[4pt]
\hat b_j^{\,s}
&\equiv
\frac{1}{\sqrt{2}}
\left(
\hat b_j^{+} + \hat b_j^{-}
\right),
&
\hat b_j^{\,a}
&\equiv
\frac{1}{\sqrt{2}}
\left(
\hat b_j^{+} - \hat b_j^{-}
\right).
\end{align}
\end{subequations}
The single-particle Hamiltonian in this new basis, which resolves the unitary symmetry that corresponds to the circulation exchange in each site, reads
\begin{align} \label{eq:H_single_sa}
    \hat{H}_{l=1}^0
    &=
    t_{s}
    \sum_{j=1}^{N_c}
    \hat{a}_j^{s\dagger}
    \hat{b}_j^{s}
    +
    t_{s}'
    \sum_{j=1}^{N_c-1}
    \hat{a}_{j+1}^{s\dagger}
    \hat{b}_j^{s}
    \nonumber\\
    &\quad
    +
    t_{a}
    \sum_{j=1}^{N_c}
    \hat{a}_j^{a\dagger}
    \hat{b}_j^{a}
    +
    t_{a}'
    \sum_{j=1}^{N_c-1}
    \hat{a}_{j+1}^{a\dagger}
    \hat{b}_j^{a}
    \nonumber\\
    &\quad
    +
    iJ_3\sin2\phi
    \sum_{j=1}^{N_c-1}
    \left(
    \hat{a}_{j+1}^{s\dagger}
    \hat{b}_{j}^{a}
    -
    \hat{a}_{j+1}^{a\dagger}
    \hat{b}_{j}^{s}
    \right)
    ,
\end{align}
with $t_s = J_2 + J_3$, $t_a = J_2 - J_3$ and $t_s' = J_2' + J_3' \cos 2\phi$, $t_a' = J_2' - J_3' \cos 2\phi$. For $\phi=0,\pi/2$, the last term in Eq.~\eqref{eq:H_single_sa} is  zero, and this Hamiltonian is decomposed into two decoupled SSH chains~\cite{Su1979Solitons}. We focus on this case in the following.
Note that, in the regime of large intercell and intracell distances $d,d'$, the coupling strengths $J_2$, $J_2'$, $J_3$, $J_3'$ are approximately equal~\cite{Pelegr2020Interaction}, and these SSH chains for $\phi=\pi/2$ are in the dimerized limit~\cite{Nicolau2023Many}.

The on-site interaction Hamiltonian $\hat{H}_{l=1}^U$ is given by
\begin{align} \label{eq:H_U}
    \hat{H}_{l=1}^U
    &=
    U
    \sum_{m=A,B}
    \sum_{j=1}^{N_c}
    \Bigg[
    \frac{1}{2}
    \hat{n}_{m,j}^{+}
    \left(
    \hat{n}_{m,j}^{+} - 1
    \right)
    \nonumber\\
    &\quad\quad\quad\quad
    +
    \frac{1}{2}
    \hat{n}_{m,j}^{-}
    \left(
    \hat{n}_{m,j}^{-} - 1
    \right)
    +
    2
    \hat{n}_{m,j}^{+}
    \hat{n}_{m,j}^{-}
    \Bigg]
    ,
\end{align}
where $U$ is the on-site interaction strength and
$\hat{n}_{m,j}^{\alpha}\equiv\hat{m}_j^{\alpha\dagger}\hat{m}_j^{\alpha}$ is the number operator of state $\ket{m_j^{\alpha}}$ for $\hat{m}=\hat{a},\hat{b}$. 
The first two terms in Eq.~\eqref{eq:H_U} correspond to the on-site interactions between atoms in states with the same circulation and have the same form as the contact interaction term in the Bose-Hubbard model~\cite{Jaksch2005The}. The last term in Eq.~\eqref{eq:H_U} accounts for the contact interactions between atoms with different circulations~\cite{nicolau2023bosonic,Nicolau2023Many,Nicolau2024Ultracold}. Note that the on-site interactions between states with different circulation acquire a factor 2 with respect to the interaction term between states with the same circulation. In the symmetric and antisymmetric basis introduced in Eq.~\eqref{eq:basis_sa_0}, the Hamiltonian in Eq.~\eqref{eq:H_U} is expressed as
\begin{align} \label{eq:Hint_sa}
    \hat{H}_{l=1}^U
    &=
    U
    \sum_{m=A,B}
    \sum_{j=1}^{N_c}
    \Bigg[
    \frac{3}{4}
    \hat{n}_{m,j}^s
    \left( \hat{n}_{m,j}^s -1
    \right)
    \nonumber\\
    &\quad
    +
    \frac{3}{4}
    \hat{n}_{m,j}^a
    \left( \hat{n}_{m,j}^a -1
    \right)
    + \hat{n}_{m,j}^s \hat{n}_{m,j}^a
    \nonumber\\
    &\quad
    -
    \frac{1}{4}
    \left(
    \hat{a}_j^{s\dagger} 
    \hat{a}_j^{s\dagger} 
    \hat{a}_j^{a}
    \hat{a}_j^{a}
    +
    \hat{a}_j^{a\dagger} 
    \hat{a}_j^{a\dagger} 
    \hat{a}_j^{s}
    \hat{a}_j^{s}
    \right)
    \nonumber\\
    &\quad
    -
    \frac{1}{4}
    \left(
    \hat{b}_j^{s\dagger} 
    \hat{b}_j^{s\dagger} 
    \hat{b}_j^{a}
    \hat{b}_j^{a}
    +
    \hat{b}_j^{a\dagger} 
    \hat{b}_j^{a\dagger} 
    \hat{b}_j^{s}
    \hat{b}_j^{s}
    \right)
    \Bigg]
    .
\end{align}
The first three terms in Eq.~\eqref{eq:Hint_sa} correspond to the contact interactions between atoms in the same or different superposition states. On the other hand, two-particle hoppings appear in the last two lines due to the difference in the interaction strengths between states with different and the same circulations. 
These interaction terms play a role only when double occupancy is allowed. As specified in the next section, however, in this work we focus on the hard-core limit where these terms are effectively removed from consideration.

The NN interaction Hamiltonian $\hat{H}_{l=1}^V$ is given by
\begin{align} \label{eq:Hnei}
    \hat{H}_{l=1}^V
    &=
    V
    \sum_{\alpha=\pm}
    \sum_{j=1}^{N_c}
    \left(
    \hat{n}_{A,j}^{\alpha} \hat{n}_{B,j}^{\alpha}
    +
    \hat{n}_{A,j}^{\alpha} 
    \hat{n}_{B,j}^{-\alpha}
    \right)
    \nonumber\\
    &\quad
    +
    V'
    \sum_{\alpha=\pm}
    \sum_{j=1}^{N_c-1}
    \left(
    \hat{n}_{B,j}^{\alpha} \hat{n}_{A,j+1}^{\alpha}
    +
    \hat{n}_{B,j}^{\alpha} 
    \hat{n}_{A,j+1}^{-\alpha}
    \right).
\end{align}
The NN interaction in our model can be induced by using dipole-dipole interactions~\cite{Baier2016Extended,Carroll2025Observation,su_dipolar_2023}. The polarization axis is oriented along the z-axis (perpendicular to the lattice xy-plane), which yields the isotropic interactions along the legs of the zig-zag ladder. The NN interaction strength can be controlled by the dipole moment and the distance between sites. Moreover, the NN interaction between $A_j$ and $A_{j+1}$ (and between $B_j$ and $B_{j+1}$) can be neglected since, for the geometry considered in our work, the same-sublattice distances between $A_j$ and $A_{j+1}$ (and between $B_j$ and $B_{j+1}$) are sufficiently larger than $d$ and $d'$.
After performing the rotation and considering the symmetric and antisymmetric states given in Eq.~\eqref{eq:basis_sa_0}, $\hat{H}_{l=1}^V$ reads
\begin{align} \label{eq:Hnei_sa}
    \hat{H}_{l=1}^V
    &=
    V
    \sum_{j=1}^{N_c}
    \sum_{\substack{\mu,\bar{\mu}=s,a\\ \bar{\mu}\neq\mu}}
    \left(
    \hat{n}_{A,j}^{\mu}
    \hat{n}_{B,j}^{\mu}
    +
    \hat{n}_{A,j}^{\mu}
    \hat{n}_{B,j}^{\bar{\mu}}
    \right)
    \nonumber\\
    &\quad
    +
    V'
    \sum_{j=1}^{N_c-1}
    \sum_{\substack{\mu,\bar{\mu}=s,a\\ \bar{\mu}\neq\mu}}
    \left(
    \hat{n}_{B,j}^{\mu} 
    \hat{n}_{A,j+1}^{\mu}
    +
    \hat{n}_{B,j}^{\mu}
    \hat{n}_{A,j+1}^{\bar{\mu}}
    \right)
    .
\end{align}
Note that the Hamiltonian in Eq.~\eqref{eq:Hnei_sa} in the symmetric and antisymmetric basis has the same structure as the Hamiltonian in Eq.~\eqref{eq:Hnei} in the original basis.

\begin{figure}[t]
    \centering
    \includegraphics[width=1.0\linewidth]{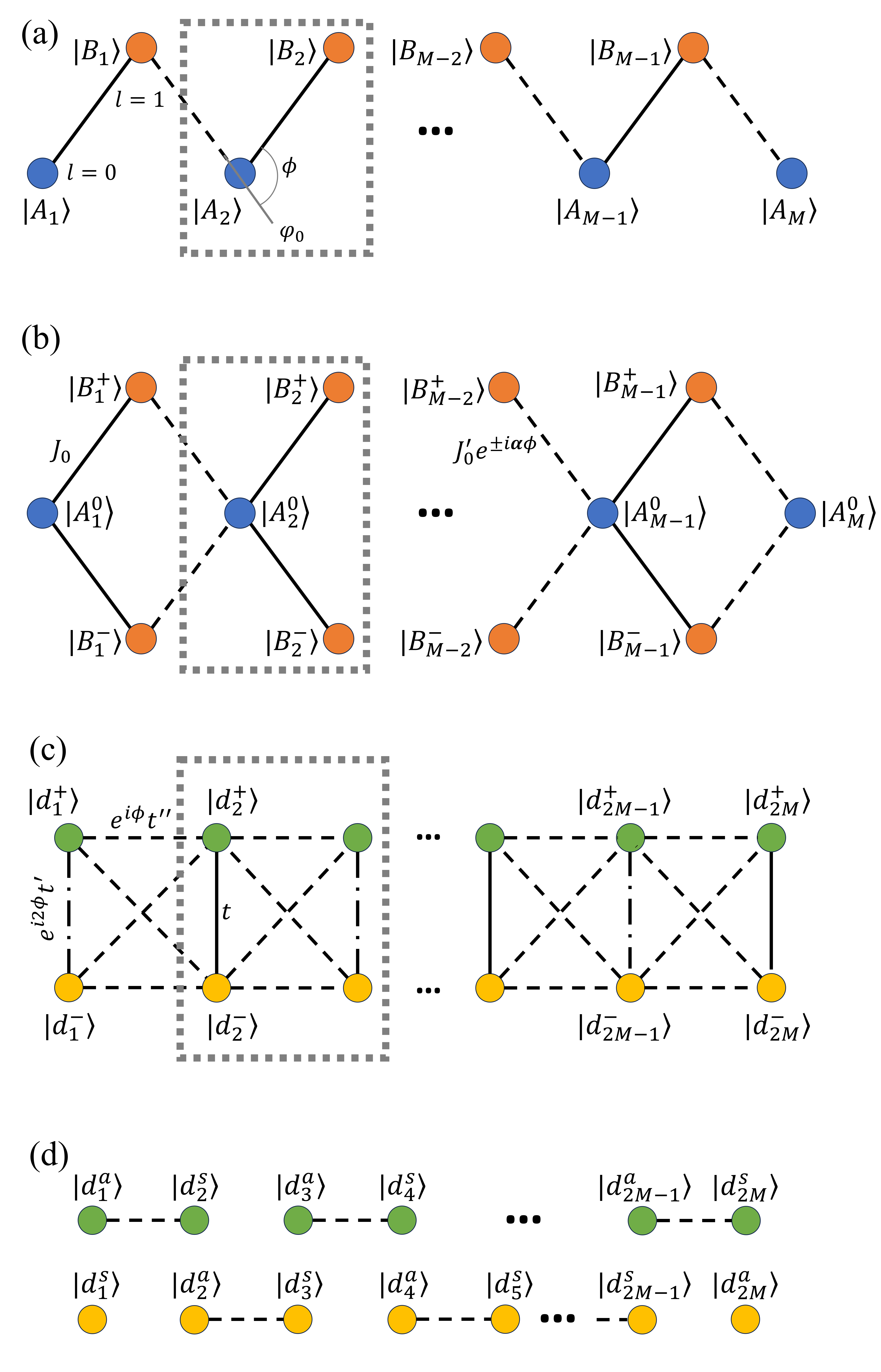}
    \caption{(a) Sketch of the 1D staggered chain of rings loaded with atoms with $l=0$ and $l=1$ in alternating sites.
    Here, the chain ends with an $A$ site, and the number of unit cells is $N_c=M+1/2$, with $M\in\mathbb{N}$. (b) Sketch of the diamond chain in the real dimension and the synthetic dimension spanned by the circulations $\pm$ in each site $B_j$.
    (c) Sketch of the extended Creutz ladder that describes the effective doublon model in Eq.~\eqref{eq:H_combo_eff}.
    The number of sites here is $2N_c-1=2M$.
    (d) Decomposition of the effective doublon model in Eq.~\eqref{eq:H_combo_eff_1_diag} into two decoupled SSH chains. The solid lines indicate real tunneling amplitudes, and the dashed and dashed-dotted lines indicate complex tunneling amplitudes between states with OAM modes with different circulations.
    The unit cells are marked by grey dotted rectangles. 
    }
    \label{fig:l01model}
\end{figure}

\subsection{Case of alternating $l=0$ and $l=1$}

Now, we consider the same setting as in Fig.~\ref{fig:l1model}(a) but let each site $A_j$ have $l=0$ and each site $B_j$ have $l=1$, so that the system forms a diamond structure in the real dimension and the synthetic dimension of circulations [see Figs.~\ref{fig:l01model}(a) and (b)]. 
The Hamiltonian in this case reads
\begin{align} \label{eq:H_combo}
   \hat{H}_{l=0,1}
   = 
   \hat{H}_{l=0,1}^0 
   +
   \hat{H}_{l=0,1}^U
   +
   \hat{H}_{l=0,1}^V
   ,
\end{align}
where
\begin{align}
    \hat{H}_{l=0,1}^0
    &=
    \sum_{\alpha=\pm}
    \left(
    J_0
    \sum_{i=1}^{N_c}
    \hat{a}_j^{0\dagger}\hat{b}_j^\alpha
    +
    J_0^{\prime}
    \sum_{j=1}^{N_c-1}
    e^{-i\alpha\phi }\hat{b}_j^{\alpha\dagger}\hat{a}_{j+1}^0
    \right)
    \nonumber\\
    &\quad
    +
    \text{H.c.},
\end{align}
with $N_c$ the number of unit cells, $J_0$ ($J_0'$) the intracell (intercell) hopping strength between OAM modes, and $e^{-i\alpha\phi}$ the complex factor for the intercell hopping between $l=0$ and the $\alpha=\pm$ circulation of $l=1$,
\begin{align}
    \hat{H}_{l=0,1}^U
    &=
    U
    \sum_{j=1}^{N_c}
    \Bigg[
    \frac{1}{2}\hat{n}_{a,j}^0(\hat{n}_{a,j}^0-1)
    \nonumber\\
    &\quad\quad
    +
    \sum_{\alpha=\pm}
    \frac{1}{2}\hat{n}_{b,j}^{\alpha}(\hat{n}_{b,j}^{\alpha} -1)
    +
    2\hat{n}_{b,j}^+\hat{n}_{b,j}^-
    \Bigg]
    ,
\end{align}
with $U$ the on-site interaction strength, and
\begin{align}
    \hat{H}_{l=0,1}^V
    =
    \sum_{\alpha=\pm}
    \left(
    V\sum_{j=1}^{N_c}
    \hat{n}_{a,j}^{0}\hat{n}_{b,j}^{\alpha}
    +
    V'\sum_{j=1}^{N_c-1}
    \hat{n}_{b,j}^{\alpha}\hat{n}_{a,j+1}^{0}
    \right)
    ,
\end{align}
with $V$ $(V')$ the intracell (intercell) NN interaction strengths.

\section{Bound states}
\label{sec:bound}

We have presented the general form of the Hamiltonian of the system in the two different layouts of rings. Now, we restrict ourselves to two particles and study bound states in the hard-core regime of infinite repulsive contact interactions ($U\to\infty$) and in the presence of large finite NN interactions compared to the hopping strengths ($V,V'\gg J_{2},J_{2}',J_{3},J_{3}'$). In this regime, two particles cannot occupy the same site. 
Moreover, the energy cost associated with having two particles at adjacent sites, $V$ or $V'$, is much higher than the one where the particles are sitting further apart, such that the former describes a tightly-bound state known as a doublon. Transitions between these two different types of two-particle states are negligible as long as $V,V'\gg J_2,J_2',J_3,J_3'$. 
We consider two particles exclusively in this work, but it is possible to generalize our discussion of doublons to clusters of $N$ particles. 

\begin{figure}[t]
    \centering
    \includegraphics[width=0.8\linewidth]{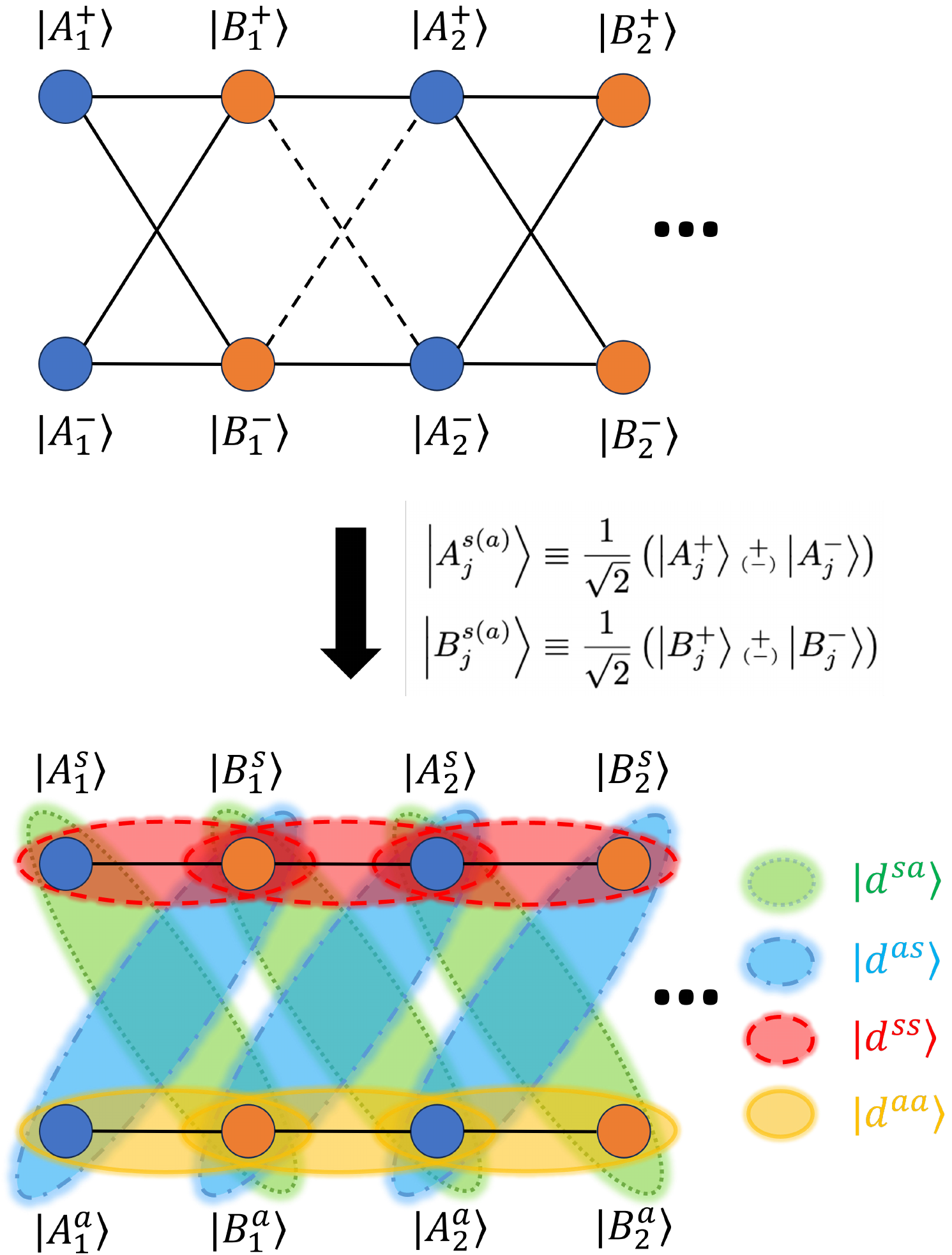}
    \caption{Description of the four types of doublons defined in Eqs.~\eqref{eq:d_ss}-\eqref{eq:d_as}.
    The red (dashed line), yellow (solid line), green (dotted line) and blue (dash-dotted line) ovals refer to Eqs.~\eqref{eq:d_ss}, ~\eqref{eq:d_aa}, ~\eqref{eq:d_sa}, and ~\eqref{eq:d_as}, respectively.
    }
    \label{fig:doublons_l1model}
\end{figure}

\subsection{Case of $l=1$} \label{sec:doublon_l1}

In the case of $l=1$, we adopt the symmetric and antisymmetric basis introduced in Eq.~\eqref{eq:basis_sa} and define the doublon states. 
There are four types of doublons in the system: both particles in the symmetric chain, both in the antisymmetric chain, or one particle in each chain.
We define the annihilation operators for the doublon in the symmetric chain as
\begin{subequations} \label{eq:d_ss}
\begin{align}
    \hat{d}_{2j-1}^{ss}
    &\equiv
    \hat{a}_{j}^{s} \hat{b}_{j}^{s}
    ,
    \\
    \hat{d}_{2j}^{ss}
    &\equiv
    \hat{b}_{j}^{s} \hat{a}_{j+1}^{s}
    ,
\end{align}
\end{subequations}
where $j$ is the unit cell index in the original system.
A bound state in one unit cell is labelled by an odd index $2j-1$, and a bound state spanning two unit cells is labelled by an even index $2j$.
The doublon in the antisymmetric chain is defined in a similar way,
\begin{subequations} \label{eq:d_aa}
\begin{align}
    \hat{d}_{2j-1}^{aa}
    &\equiv
    \hat{a}_{j}^{a} \hat{b}_{j}^{a}
    ,
    \\
    \hat{d}_{2j}^{aa}
    &\equiv
    \hat{b}_{j}^{a} \hat{a}_{j+1}^{a}
    ,
\end{align}
\end{subequations}
and the doublons bridging the two chains are defined as
\begin{subequations} \label{eq:d_sa}
\begin{align}
    \hat{d}_{2j-1}^{sa}
    &\equiv
    \hat{a}_{j}^{s} \hat{b}_{j}^{a}
    ,
    \\
    \hat{d}_{2j}^{sa}
    &\equiv
    \hat{b}_{j}^{s} \hat{a}_{j+1}^{a}
    ,
\end{align}
\end{subequations}
and
\begin{subequations} \label{eq:d_as}
\begin{align}
    \hat{d}_{2j-1}^{as}
    &\equiv
    \hat{a}_{j}^{a} \hat{b}_{j}^{s}
    ,
    \\
    \hat{d}_{2j}^{as}
    &\equiv
    \hat{b}_{j}^{a} \hat{a}_{j+1}^{s}
    .
\end{align}
\end{subequations}
These four types of doublons are depicted in Fig.~\ref{fig:doublons_l1model}.

\begin{figure}[t]
    \includegraphics[width=0.9\linewidth]{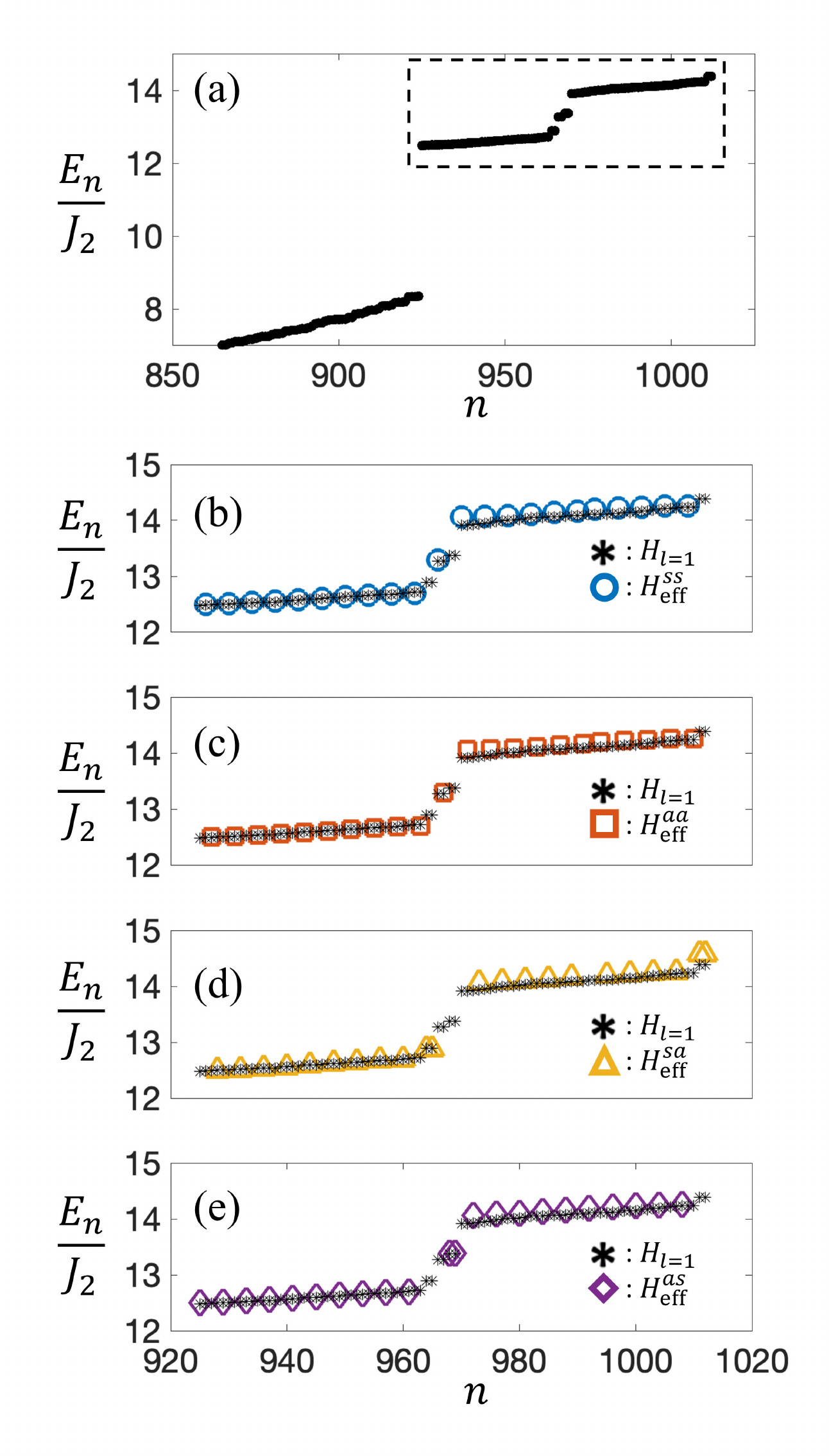}
    \caption{(a) Energy spectrum of the original Hamiltonian $\hat{H}_{l=1}$ in Eq.~\eqref{eq:H_total}, zoomed into the energy window close to the doublon subspace.
    (b, c, d, e) Eigenvalues of the different terms of the effective Hamiltonian, such as (b) $\hat{H}_{\text{eff}}^{ss}$ and (c) $\hat{H}_{\text{eff}}^{aa}$ in Eq.~\eqref{eq:H_ss_aa} and (d) $\hat{H}_{\text{eff}}^{sa}$ and (e) $\hat{H}_{\text{eff}}^{as}$ in Eq.~\eqref{eq:H_sa_as}, within the region of the spectrum corresponding to the doublon subspace [dashed rectangle (a)]. Parameter values: $J_3/J_2=2.125$, $J_2'=J_2$, $J_3'=J_3$, $\phi=\pi/2$, $V/J_2=12.5$, and $N_{c}=11.5$.
     }
    \label{fig:doublon_l1}
\end{figure}

We expand the single-particle Hamiltonian in Eq.~\eqref{eq:H_single_sa} in the manifold of doublons and introduce the hopping terms $t_s$ and $t_a$ as a perturbation. To simplify the calculations, we restrict ourselves to the particular case $V=V'$.
Assuming $V\gg J_2,J_2',J_3,J_3'$ as well as $U\to\infty$ and using perturbation theory up to second-order corrections, the following effective Hamiltonian, consisting of four decoupled chains, can be derived (see Appendix~\ref{app:H_eff_l1} for details),
\begin{align} \label{eq:H_eff_l1}
    \hat{H}_{l=1}^{\text{eff}}
    &=
    \hat{H}_{\text{eff}}^{ss}\oplus \hat{H}_{\text{eff}}^{aa}\oplus \hat{H}_{\text{eff}}^{sa}\oplus \hat{H}_{\text{eff}}^{as}
    ,
\end{align}
where
\begin{align}
    \label{eq:H_ss_aa}
    \hat{H}_{\text{eff}}^{\mu\mu}
    &=
    \sum_{r=1}^{2N_c-1}
    \epsilon_{r}^{\mu\mu}
    \hat{d}_{r}^{\mu\mu\dagger} \hat{d}_{r}^{\mu\mu}
    +
    \left(
    \frac{t_{\mu}'t_{\mu}}{V}
    \sum_{r=1}^{2N_c-2}
    \hat{d}_{r+1}^{\mu\mu\dagger} \hat{d}_{r}^{\mu\mu}
    +\text{H.c.}
    \right)
    ,
\end{align}
with
\begin{align}
\begin{cases}
    \epsilon_1 ^{\mu\mu}
    =
    V + t_{\mu}'^2/V,
    \\
    \epsilon_{2n}^{\mu\mu}
    =
    V + 2t_{\mu}^2/V 
    &
    \text{for } 
    2\leq 2n \leq 2N_c-2,
    \\
    \epsilon_{2n+1}^{\mu\mu}
    =
    V + 2t_{\mu}'^2/V 
    &
    \text{for } 
    3\leq 2n+1 \leq 2N_c-2,
    \\
    \epsilon_{2N_c-1}^{\mu\mu}
    =
    V+t_{\mu}'^2/V
    & 
    \text{if $N_{c}$ is integer},
    \\
    \epsilon_{2N_c-1}^{\mu\mu}
    =
    V+t_{\mu}^2/V
    & 
    \text{if $N_{c}$ is half-integer}
    ,
\end{cases}
\end{align}
and
\begin{align}
    \label{eq:H_sa_as}
    \hat{H}_{\text{eff}}^{\mu\bar{\mu}}
    &=
    \sum_{j=1}^{2N_{c}-1}
    \epsilon_{j}^{\mu\bar{\mu}}
    \hat{d}_{j}^{\mu\bar{\mu}\dagger} \hat{d}_{j}^{\mu\bar{\mu}}
    +
    \bigg(
    \frac{t_{\bar{\mu}}'t_{\mu}}{V}
    \sum_{j=1}^{\floor*{N_{c}-1/2}}
    \hat{d}_{2j}^{\mu\bar{\mu}\dagger} \hat{d}_{2j-1}^{\mu\bar{\mu}}
    \nonumber\\
    &\quad
    +
    \frac{t_{\mu}'t_{\bar{\mu}}}{V}
    \sum_{j=1}^{\floor*{N_{c}-1}}
    \hat{d}_{2j+1}^{\mu\bar{\mu}\dagger} \hat{d}_{2j}^{\mu\bar{\mu}}
    +\text{H.c.}
    \bigg)
    ,
\end{align}
with
\begin{align} 
\label{eq:epsilon_sa_as}
\begin{cases}
    \epsilon_1 ^{\mu\bar{\mu}}
    =
    V + t_{\bar{\mu}}'^2/V,
    \\
    \epsilon_{2j}^{\mu\bar{\mu}}
    =
    V + (t_{\mu}^2+t_{\bar{\mu}}^2)/V 
    &
    \text{for } 
    2\leq 2j \leq 2N_c-2,
    \\
    \epsilon_{2j+1}^{\mu\bar{\mu}}
    =
    V + (t_{\mu}'^2+t_{\bar{\mu}}'^2)/V 
    &
    \text{for } 
    3\leq 2j+1 \leq 2N_c-2,
    \\
    \epsilon_{2N_c-1}^{\mu\bar{\mu}}
    =
    V+t_{\mu}'^2/V
    &
    \text{if $N_c$ is integer},
    \\
    \epsilon_{2N_c-1}^{\mu\bar{\mu}}
    =
    V+t_{\mu}^2/V
    &
    \text{if $N_c$ is half-integer},
\end{cases}
\end{align}
for $\mu,\bar{\mu}=s,a$ and $\bar{\mu}\neq\mu$. The sketch of $\hat{H}_{l=1}^{\text{eff}}$ in Eq.~\eqref{eq:H_eff_l1} is given in Fig.~\ref{fig:l1model}(c).
The function $\floor*{x}$ is the floor function, which gives the highest integer less than or equal to $x$.
The number of sites in each chain is $2N_c-1$. Note that the cell number $N_c$ can be an integer but also a half integer, and the latter means that the chain ends with an $A$ site. While the chains described by $\hat{H}_{\text{eff}}^{ss}$ and $\hat{H}_{\text{eff}}^{aa}$ in Eq.~\eqref{eq:H_ss_aa} have uniform hopping strength, the chains $\hat{H}_{\text{eff}}^{sa}$ and $\hat{H}_{\text{eff}}^{as}$ in Eq.~\eqref{eq:H_sa_as} have two alternating hopping strengths similar to the SSH chain. 
In all the chains, non-uniform effective on-site energies appear, which break the chiral symmetry. Nevertheless, this on-site energy mismatch can be compensated by tuning the parameters in the original Hamiltonian $\hat{H}_{l=1}$ in Eq.~\eqref{eq:H_total}~\cite{nicolau2023bosonic,Nicolau2023Many}. As an example, we recover the chiral symmetry in $\hat{H}_{\text{eff}}^{as}$ in Eq.~\eqref{eq:H_sa_as}.
The effective on-site energies at odd indices and even indices are equal if $t_s^2+t_a^2=t_s'^2+t_a'^2$. A simple option to satisfy this condition is $J_2=J_2'$ and $J_3=J_3'$, which can be achieved by fixing $d=d'$. In addition, the effective on-site energy of doublon states occupying an edge site is smaller than that of those occupying bulk sites only, since the lower coordination number of the edge sites translates into one less mediating state available for the doublon states that occupy them. 
Such mismatch can be modified by tuning the NN interaction $V$ at the edges. Denoting the interaction strength at the left edge with $V_L$ and using Eq.~\eqref{eq:epsilon_sa_as}, the compensation reads
\begin{align}
    V_L
    &=
    V
    +
    \epsilon_2^{as}
    -
    \epsilon_1^{as}
    \nonumber\\
    &=
    V
    +
    \frac{t_s^2+t_a^2-t_s'^2}{V}
    ,
\end{align}
which makes the on-site energy uniform up to the second-order. 
Notice that this change not only corrects the on-site energy but also affects the effective hopping at the left edge. However, the influence is negligible as
\begin{align}
    \frac{t_s't_a}{V_L}
    &\approx
    \frac{t_s't_a}{V}
    \left(
    1-
    \frac{t_s^2+t_a^2-t_s'^2}{V^2}
    \right)
    .
\end{align}
The compensation for the right edge is obtained in the same way and is given by $V_R=V+\epsilon_{2N_c-1}^{as}-\epsilon_{2N_c-2}^{as}$.
Those changes are for the chain $\hat{H}_{\text{eff}}^{as}$, and one cannot compensate the other chain $\hat{H}_{\text{eff}}^{sa}$ simultaneously. 
With such correction, the bulk-edge correspondence is recovered, and $\hat{H}_{\text{eff}}^{as}$ corresponds to an SSH chain.
Considering $N_c$ half-integer, the number of sites in $\hat{H}_{\text{eff}}^{as}$ is even [see Fig.~\ref{fig:l1model}(c)].
In this case, two topological states appear at both edges when the ratio $t_a't_s/t_s't_a$ is larger than one. 
If $N_c$ is an integer and the number of sites in $\hat{H}_{\text{eff}}^{as}$ is odd, one topological state emerges at the left or right edge, depending on the ratio $t_a't_s/t_s't_a$.

Figure~\ref{fig:doublon_l1}(a) shows the energy spectrum of $\hat{H}_{l=1}$ in Eq.~\eqref{eq:H_total} in the energy regime close to the normalised doublon self-energy $V/J_2$.
It is obtained by using exact diagonalization for $N_c=11.5$, $J_3/J_2=2.125$, $J_2'=J_2$, $J_3'=J_3$, $\phi=\pi/2$, and $V/J_2=12.5$ so that the ratio is $t_a't_s/t_s't_a>1$. The spectrum can be divided into two sectors: the lower continuum of states corresponding to states of two (approximately) non-interacting particles, and the higher-energy states in the doublon subspace corresponding to a heavy particle and composed of two gapped bands with four in-gap states and two states above the top bands as shown in Figs.~\ref{fig:doublon_l1}(b)-(e).
Even when the ratios $V/J_2$ and $V/J_3$ are still away from the strong interactions regime ($V/J_2=12.5$, $V/J_3=5.88$), the energy spectrum of the effective Hamiltonian $\hat{H}_{l=1}^{\text{eff}}$ in Eq.~\eqref{eq:H_eff_l1} shows a good agreement with that of the original Hamiltonian $\hat{H}_{l=1}$ in Eq.~\eqref{eq:H_total} [see Figs.~\ref{fig:doublon_l1}(b)-(e)].
There are six in-gap states, and the two with the highest energy correspond to the edge states of the compensated $\hat{H}_{\text{eff}}^{as}$ [diamonds in Fig.~\ref{fig:doublon_l1}(e)]. The other in-gap states arise from the non-uniform on-site energy at the edge sites of $\hat{H}_{\text{eff}}^{ss}$, $\hat{H}_{\text{eff}}^{aa}$, and $\hat{H}_{\text{eff}}^{sa}$.
There are small deviations between the effective Hamiltonian and the original one in the upper band. We have checked that the deviation vanishes with increasing $V$. Furthermore, the two highest energy states indicated with yellow triangles in Fig.~\ref{fig:doublon_l1}(d) show a larger mismatch between the effective chain $\hat{H}_{\text{eff}}^{sa}$ and the original one. This comes from the compensation of the on-site energies at the edges for the effective chain $\hat{H}_{\text{eff}}^{as}$ by means of the interaction strengths $V_L,V_R$. 
\begin{figure}[bt]
    \includegraphics[width=0.7\linewidth]{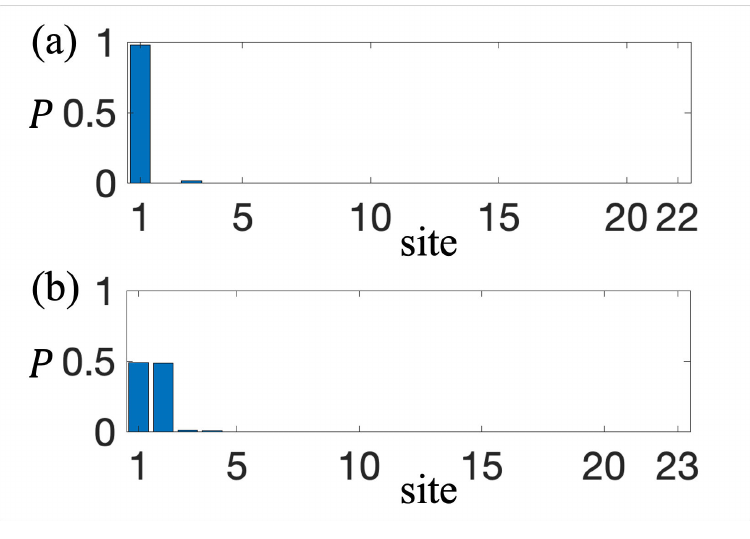}
    \caption {Normalized density profile of the topological left edge state in (a) the effective model $\hat{H}_{\text{eff}}^{as}$ and in (b) the original lattice. 
    The same parameter values as in Fig.~\ref{fig:doublon_l1} are used. 
    In (a), the odd (even) sites correspond to $\ket{d_{2j-1}^{as}}$ ($\ket{d_{2j}^{as}}$). The number of sites is $2N_c-1=22$.
    In (b), the odd (even) sites correspond to sites $A_j$ ($B_j$). The number of sites is $2N_c=23$. 
    }
    \label{fig:density_l1}    
\end{figure}

Let us now focus on the topological edge states of the effective chain $\hat{H}_{\text{eff}}^{as}$. The density profile of the topological doublon left edge state is shown in Fig.~\ref{fig:density_l1}(a). The first and third sites are populated while the second site is empty, i.e. only the sublattice of odd sites is occupied. We convert the basis of the effective doublon chain $\hat{H}_{\text{eff}}^{as}$ into the basis of the single particle system and display the population at sites $A_j$ and $B_j$ in the $j$th unit cell of the original model [see Fig.~\ref{fig:density_l1}(b)]. In contrast to panel~(a), the first and the second sites, $A_1$ and $B_1$, are both highly populated, and the third and the fourth sites, $A_2$ and $B_2$, have small but non-negligible population. This is because the population of the doublon on the first (third) site in Fig.~\ref{fig:density_l1}(a) has contribution of sites $A_{1},B_{1}$ ($A_{2},B_{2}$) in the original system in Fig.~\ref{fig:density_l1}(b).

\subsection{Case of alternating $l=0$ and $l=1$} \label{sec:doublon_l01}

\begin{figure}[t]
    \centering
    \includegraphics[width=0.9\linewidth]{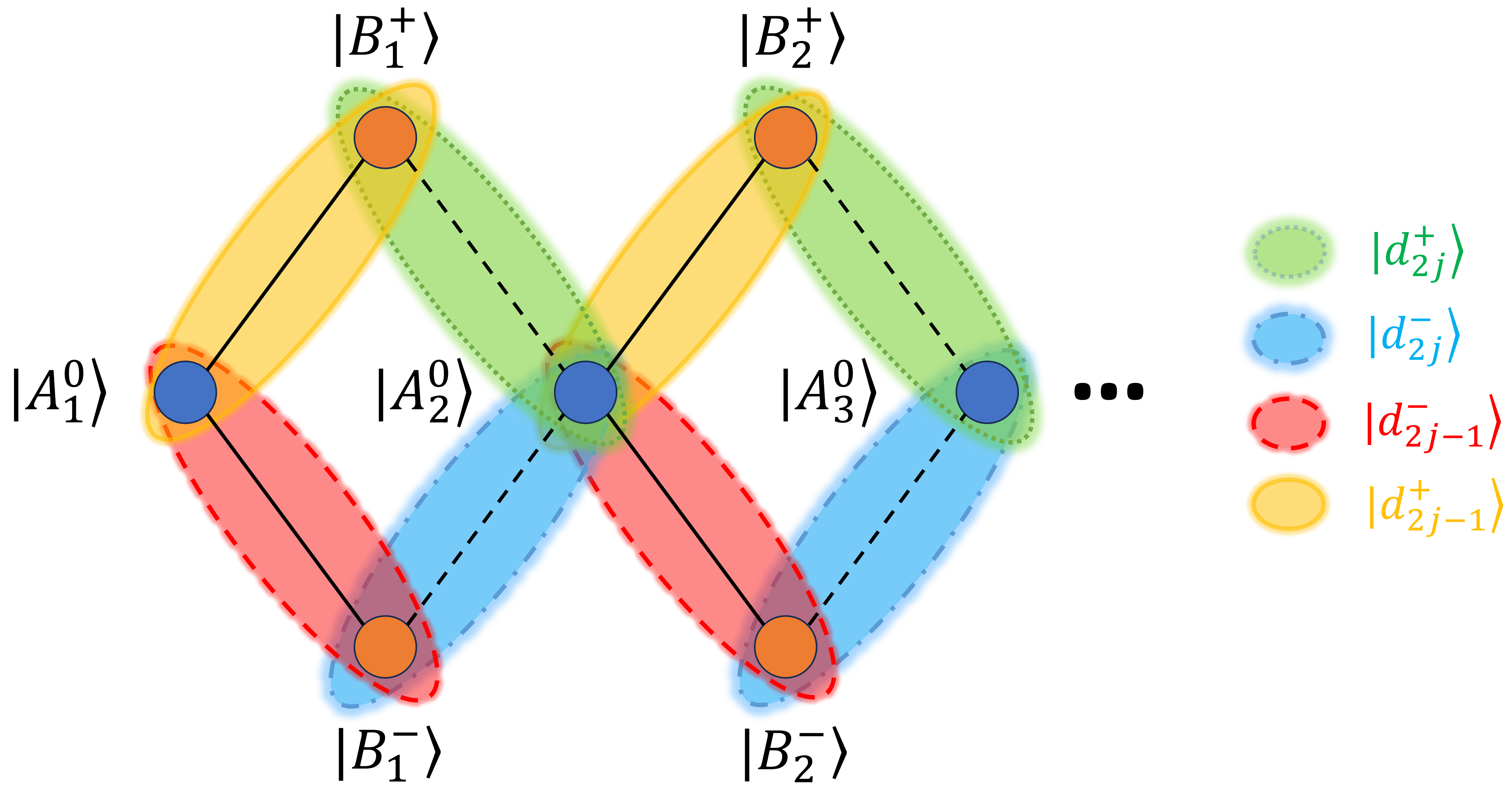}
    \caption{Description of the four types of doublons defined in Eq.~\eqref{eq:doublons_diamond}.
    The yellow (solid line), red (dashed line), green (dotted line) and blue (dash-dotted line) ovals refer to Eqs.~\eqref{eq:doublons_diamond_plus_odd},~\eqref{eq:doublons_diamond_minus_odd},~\eqref{eq:doublons_diamond_plus_even},~\eqref{eq:doublons_diamond_minus_even}, respectively.
    }
    \label{fig:doublons_l01model}
\end{figure}

Let us now consider the system in Fig.~\ref{fig:l01model}(a).
As in the previous subsection, we take infinite repulsive on-site interaction, $U\to\infty$, and impose $V\gg J_0,J'_0$ so that two particles sitting next to each other are bound. We start by imposing uniform NN interaction $V=V'$, although we take advantage of $V\neq V'$ later. In this system, there are also four possible bound states with annihilation operators
\begin{subequations} \label{eq:doublons_diamond}
\begin{align}
    \hat{d}_{2j-1}^{+}
    &\equiv 
    \hat{a}_j^{0}\hat{b}_j^{+}
    ,
    \label{eq:doublons_diamond_plus_odd}
    \\
    \hat{d}_{2j-1}^{-}
    &\equiv 
    \hat{a}_j^{0}\hat{b}_j^{-}
    ,
    \label{eq:doublons_diamond_minus_odd}
    \\
    \hat{d}_{2j}^{+}
    &\equiv 
    \hat{b}_j^{+}\hat{a}_{j+1}^{0}
    ,
    \label{eq:doublons_diamond_plus_even}
    \\
    \hat{d}_{2j}^{-}
    &\equiv 
    \hat{b}_j^{-}\hat{a}_{j+1}^{0}
    .
    \label{eq:doublons_diamond_minus_even}
\end{align}
\end{subequations}
These four types of doublons are depicted in Fig.~\ref{fig:doublons_l01model}.

\begin{figure}
    \centering
    \includegraphics[width=0.6\linewidth]{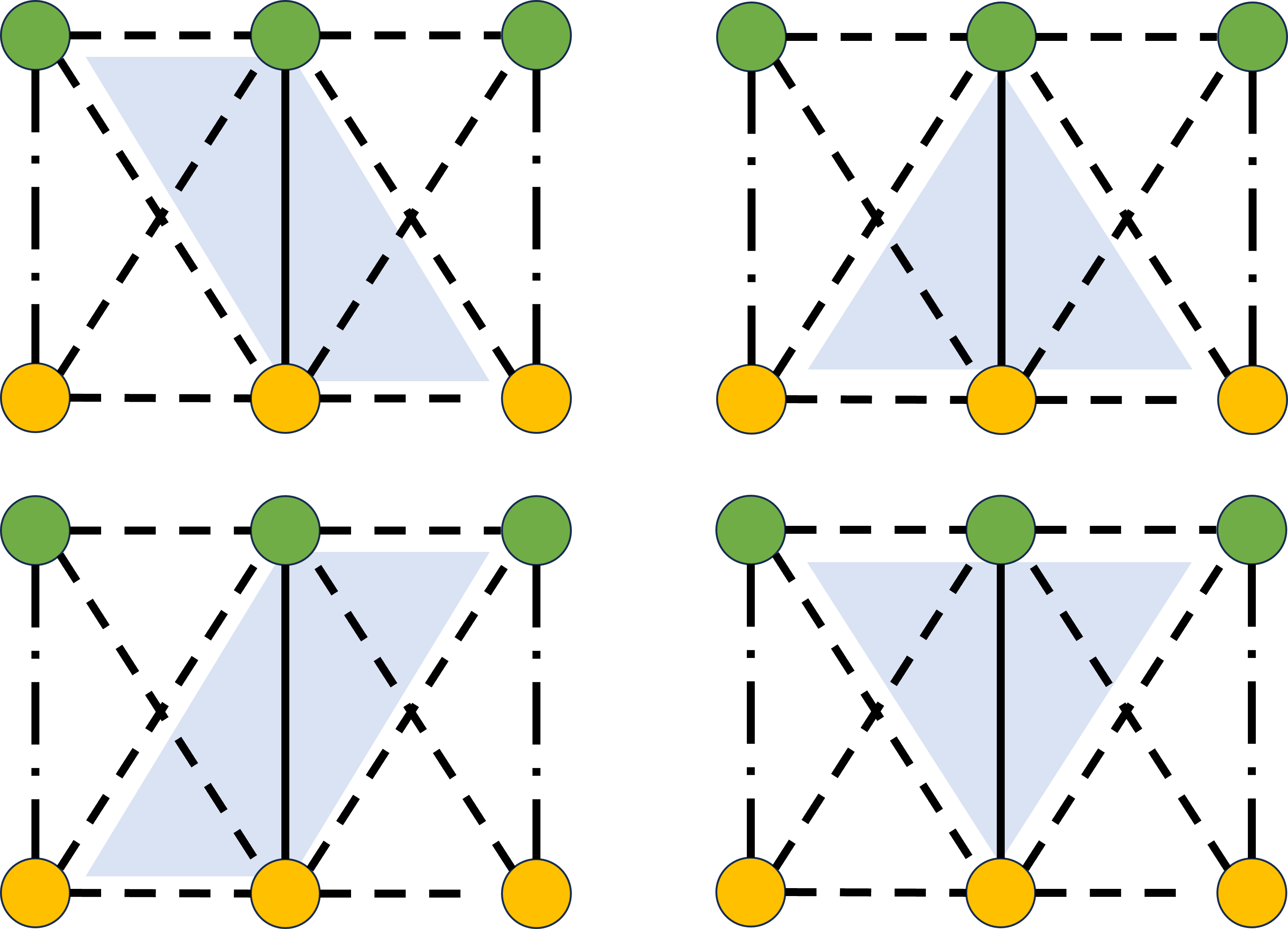}
    \caption{Sketch of a portion of the ladder depicted in Fig.~\ref{fig:l01model}(c) with $\phi=\pi/2$ fixed. The shaded regions in the diagrams correspond to the different loops that enclose a $\pi$-flux.}
    \label{fig:piflux}
     \vspace{-6pt}
\end{figure}

By considering $V=V'\gg J_0,J'_0$ and using perturbation theory up to second-order corrections, the effective Hamiltonian for the doublons can be derived (the details are given in Appendix~\ref{app:H_eff_l1}), corresponding to a Creutz ladder with extra vertical hoppings,
\begin{align} \label{eq:H_combo_eff}
    \hat{H}_{l=0,1}^{\text{eff}}
    &=
    \sum_{j=1}^{2N_c-1} 
    \epsilon_j
    \left(
    \hat{d}_{j}^{+\dagger} 
    \hat{d}_{j}^{+}
    +
    \hat{d}_{j}^{-\dagger} 
    \hat{d}_{j}^{-}
    \right)
    \nonumber\\
    &\quad
    +
    \bigg(
    e^{i\phi}
    t''
    \sum_{j=1}^{2N_c-2}
    \left[
    \hat{d}_{j+1}^{+\dagger} \hat{d}_{j}^{+} 
    - 
    \hat{d}_{j+1}^{-\dagger} 
    \hat{d}_{j}^{-} 
    \right]
    \nonumber\\
    &\quad
    +
    e^{2i\phi}
    t'
    \sum_{j=1}^{\floor{N_c}} 
    \hat{d}_{2j-1}^{-\dagger} 
    \hat{d}_{2j-1}^{+}
    +
    t
    \sum_{j=1}^{\floor*{N_c-1/2}} 
    \hat{d}_{2j}^{-\dagger} 
    \hat{d}_{2j}^{+}
    \nonumber\\
    &\quad
    +
    e^{i\phi}
    t''
    \sum_{j=1}^{2N_c-2}
    \left[
    -
    \hat{d}_{j+1}^{+\dagger} \hat{d}_{j}^{-}
    +
    \hat{d}_{j+1}^{-\dagger} \hat{d}_{j}^{+}
    \right]
    +
    \text{H.c.}
    \bigg)
    ,
\end{align}
where $t\equiv J_0^2/V$, $t'\equiv J_0'^2/V$, $t''\equiv J_0J_0'/V=\sqrt{tt'}$, and
\begin{align} \label{eq:epsilon_combo}
\begin{cases}
    \epsilon_1
    =
    V + J_0'^2/V
    ,
    \\
    \epsilon_{2j}
    =
    V + 3J_0^2/V
    ,
    &
    \text{for } 
    2\leq 2j \leq M-1
    ,
    \\
    \epsilon_{2j+1}
    =
    V + 3J_0'^2/V
    ,
    &
    \text{for } 
    3\leq 2j+1 \leq M-1
    ,
    \\
    \epsilon_{M}
    =
    V + J_0^2/V
    .
\end{cases}
\end{align}
The first line in Eq.~\eqref{eq:H_combo_eff} gives the on-site energies. 
The second line represents the intra-leg hopping, and the third line shows inter-leg hopping along vertical rungs.
The fourth line shows diagonal links [see Fig.~\ref{fig:l01model}(c)]. 
The unit cell is extended into four sites. 

As in the previous section, we can compensate for the bulk-edge potential mismatch to recover the bulk-edge correspondence.
According to Eq.~\eqref{eq:epsilon_combo}, the on-site energy in the bulk is uniform when $J_0=J_0'$, and the on-site energy at the edges is compensated by increasing the NN interaction at the edges. Denoting the NN interaction strengths at the left and right edges with $V_L$ and $V_{R}$, respectively, the on-site energy is made uniform by the change $V_L=V+\epsilon_2-\epsilon_1$ and $V_R=V+\epsilon_{M-1}-\epsilon_{M}$.

While the staggering angle $\phi$ can be tuned in the interval $[0,2\pi/3)$, the case of $\phi=\pi/2$ is particularly interesting as a $\pi$-flux threads each plaquette (see Fig.~\ref{fig:piflux}). This scenario is not seen in the single-particle case and is unique to the doublon regime~\cite{Croft2023Anomalous}. 
Therefore, we focus on the case of $\phi=\pi/2$. Let us rewrite $\hat{H}_{l=0,1}^{\text{eff}}$ in Eq.~\eqref{eq:H_combo_eff} after correcting for the on-site energy mismatch at the end sites for the case of $J_0=J_0'$, 
\begin{align} \label{eq:H_combo_eff_1}    
    \hat{H}_{l=0,1}^{\text{eff,}1}
    &= 
    it
    \sum_{j=1}^{2N_c-2} 
    \left[
    \hat{d}_{j+1}^{+\dagger} \hat{d}_{j}^{+} 
    - 
    \hat{d}_{j+1}^{-\dagger} 
    \hat{d}_{j}^{-} 
    \right]
    \nonumber\\
    &\quad
    -
    t
    \sum_{j=1}^{\floor*{N_c}} 
    \hat{d}_{2j-1}^{-\dagger} 
    \hat{d}_{2j-1}^{+}
    +
    t
    \sum_{j=1}^{\floor*{N_c-1/2}} 
    \hat{d}_{2j}^{-\dagger} 
    \hat{d}_{2j}^{+}
    \nonumber\\
    &\quad
    +
    it
    \sum_{j=1}^{2N_c-2}
    \left[
    -
    \hat{d}_{j+1}^{+\dagger} \hat{d}_{j}^{-}
    +
    \hat{d}_{j+1}^{-\dagger} \hat{d}_{j}^{+}
    \right]
    +
    \text{H.c.}
    ,
\end{align}
where the featureless uniform on-site energy term was dropped for convenience. 

\begin{figure}[tb]
    \includegraphics[width=0.9\linewidth]{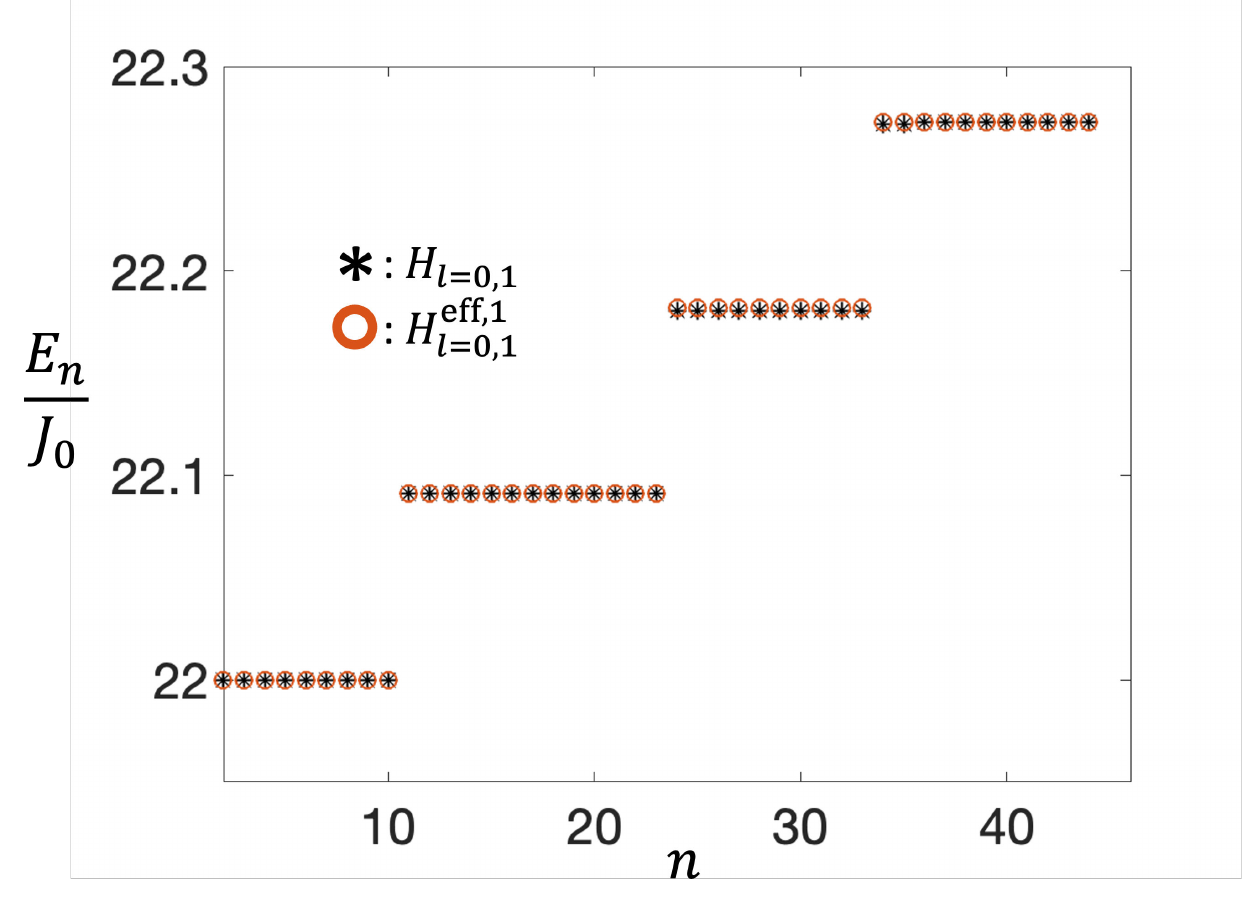}
    \caption{Energy spectrum window of the original Hamiltonian $\hat{H}_{l=0,1}$ in Eq.~\eqref{eq:H_combo} corresponding to the doublons and of the effective doublon Hamiltonian $\hat{H}_{l=0,1}^{\text{eff,}1}$ in Eq.~\eqref{eq:H_combo_eff_2} for $J_0=J_0^\prime$, $\phi=\pi/2$, 
    $V/J_0=V'/J_0=22$,
    and $N_{c}=11.5$. 
    }
    \label{fig:spectrum_CL}
\end{figure}

There can be two different right ends for the chain $\hat{H}_{l=0,1}^{\text{eff,}1}$ in Eq.~\eqref{eq:H_combo_eff_1}: (i) the last site is an $A$ site [see Figs.~\ref{fig:l01model}(a),(b)] and (ii) the last site is a $B$ site [see Figs.~\ref{fig:diamond_broken}(a),(b)], corresponding to half-integer $N_c$ and integer $N_c$, respectively.
We will consider the former case here and discuss the latter in the Appendix~\ref{app:bsite_ending}.

\subsubsection{A site ending}
If the original chain ends with an $A$ site, as shown in Fig.~\ref{fig:l01model}(b), all the sites in the effective chain $\hat{H}_{l=0,1}^{\text{eff,}1}$ in Eq.~\eqref{eq:H_combo_eff_1} are coupled through the effective inter-leg hoppings [see Fig.~\ref{fig:l01model}(c)]
, and 
in this case we have $N_c=M+1/2$ with $M\in\mathbb{N}$. The presence of a $\pi$-flux in the plaquette of the effective Creutz model is related to the existence of compact localized states (CLSs)~\cite{Bodyfelt2014,Maimaiti2017}.
We define the annihilation operators of the CLSs as ~\cite{Kuno2020Extended,Tovmasyan2018thesis}
\begin{align} \label{eq:op_CLS}
    \hat{W}_{j,\pm}
    &\equiv
    \pm\frac i2 \hat{d}_{j}^{+}
    \mp\frac i2 \hat{d}_{j}^{-}
    +\frac12 \hat{d}_{j+1}^{+}
    +\frac12 \hat{d}_{j+1}^{-}
    ,
\end{align}
which satisfy the bosonic commutation relation $[\hat{W}_{j,\alpha},\hat{W}_{k,\alpha'}^{\dagger}]=\delta_{j,k}\delta_{\alpha,\alpha'}$. In the same way, the annihilation operators for the edge states read, respectively, as~\cite{pelegri2024few} 
\begin{subequations} \label{eq:op_LR}
\begin{align}
    \hat{L}
    &\equiv
    \frac{1}{\sqrt{2}}
    \left(
    \hat d_{1}^{+}
    +
    \hat d_{1}^{-}
    \right)
    ,
    \\
    \hat{R}
    &\equiv
    \frac{1}{\sqrt{2}}
    \left(
    \hat d_{2M}^{+}
    -
    \hat d_{2M}^{-}
    \right)
    .
\end{align}
\end{subequations}
The operators in Eqs.~\eqref{eq:op_CLS} and \eqref{eq:op_LR} diagonalise $\hat{H}_{l=0,1}^{\text{eff,}1}$ in Eq.~\eqref{eq:H_combo_eff_1} as~\cite{Kuno2020Extended}
\begin{align} \label{eq:H_combo_eff_2}
    \hat{H}_{l=0,1}^{\text{eff,}1}
    &=
    \sum_{j=1}^{2M-1} 
    \left[
    t
    \left(
    2-(-1)^j
    \right)
    \hat{W}_{j,+}^{\dagger} \hat{W}_{j,+}
    \right.
    \nonumber\\
    &\quad
    \left.
    +
    t
    \left(
    -2-(-1)^j
    \right)
    \hat{W}_{j,-}^{\dagger} \hat{W}_{j,-}
    \right]
    \nonumber\\
    &\quad
    -t
    \left(
    \hat{L}^{\dagger}\hat{L} 
    +
    \hat{R}^{\dagger}\hat{R}
    \right)
    .
\end{align}
The first two lines indicate that there are four flat bands with energies $-3t,-t,t,3t$. 
The energy of the edge states $\ket{L},\ket{R}$ is $-t$, and they are contained in one of the bands.
We have computed the energy spectrum of the original Hamiltonian $\hat{H}_{l=0,1}$ in Eq.~\eqref{eq:H_combo} and observed good agreement with that of the effective Hamiltonian in Eq.~\eqref{eq:H_combo_eff_2} (see Fig.~\ref{fig:spectrum_CL}).

To characterise topologically the obtained edge states, we employ the symmetric and antisymmetric basis with annihilation operators
\begin{align}
\label{eq:op_d_sa}
\hat{d}_{j}^{s}
&\equiv
\frac{1}{\sqrt{2}}
\left(
\hat{d}_{j}^{+} + \hat{d}_{j}^{-}
\right),
&
\hat{d}_{j}^{a}
&\equiv
\frac{1}{\sqrt{2}}
\left(
\hat{d}_{j}^{+} - \hat{d}_{j}^{-}
\right).
\end{align}
In this basis, $\hat{H}_{l=0,1}^{\text{eff,}1}$ in Eq.~\eqref{eq:H_combo_eff_1} is expressed as
\begin{align} \label{eq:H_combo_eff_1_diag}
    \hat{H}_{l=0,1}^{\text{eff,}1}
    &=
    \sum_{j=1}^{M}
    \bigg[
    \left(
    -2it
    \hat{d}_{2j-1}^{a\dagger} 
    \hat{d}_{2j}^{s}
    +
    \text{H.c.}
    \right)
    \nonumber\\
    &\quad\quad
    +t
    \left(
    \hat{d}_{2j-1}^{a\dagger}
    \hat{d}_{2j-1}^{a}
    +
    \hat{d}_{2j}^{s\dagger}
    \hat{d}_{2j}^{s}
    \right)
    \bigg]
    \nonumber\\
    &
    +
    \sum_{j=1}^{M-1}
    \bigg[
    \left(
    -2it
    \hat{d}_{2j}^{a\dagger} 
    \hat{d}_{2j+1}^{s}
    +
    \text{H.c.}
    \right)
    \nonumber\\
    &\quad\quad
    -t
    \left(
    \hat{d}_{2j}^{a\dagger}
    \hat{d}_{2j}^{a}
    +
    \hat{d}_{2j+1}^{s\dagger}
    \hat{d}_{2j+1}^{s}
    \right)
    \bigg]
    \nonumber\\
    &-t
    \left(
    \hat{d}_1^{s\dagger}\hat{d}_1^{s}
    +
    \hat{d}_{2M}^{a\dagger}\hat{d}_{2M}^{a}
    \right)
    ,
\end{align}
which describes two decoupled SSH chains in the dimerized limit. 
The first two lines correspond to an SSH chain in the trivial phase, and the rest corresponds to an SSH chain in the topological phase, as depicted in Fig.~\ref{fig:l01model}(d). 
Therefore, there are two topological edges states $\ket{d_1^s},\ket{d_{2M}^a}$, which correspond to $\ket{L}$ and $\ket{R}$ defined through Eqs.~\eqref{eq:op_LR}, respectively.
We emphasise that, even though every coupling strength in the original chain is equal, this system nevertheless acquires doublon topological states embedded in one of the flat bands of the effective system.

\subsubsection{Varying vertical coupling}

Up to this point, we have considered $V=V'$ and $J_0=J_0'$ so that the on-site energy in the bulk of the effective doublon model is uniform. 
Let us now consider the case of $V\neq V'$ and $J_0\neq J_0'$ and impose $V'-V=3(J_0'^2-J_0^2)/V$ in the bulk to make the bulk on-site energy uniform as $\epsilon_{2j+1}=\epsilon_{2j}$ [see Eq.~\eqref{eq:epsilon_combo}]. This allows us to freely tune the hopping ratio $t/t'$ of the effective Hamiltonian $\hat{H}_{l=0,1}^{\text{eff}}$ in Eq.~\eqref{eq:H_combo_eff}. We take into account $t\neq t'$ and introduce the following CLSs annihilation operators
\begin{align} \label{eq:op_W_dd}
    \hat{W}''_{j,\pm}
    &\equiv
    \frac{1}{\sqrt{2(1+\gamma_{\pm}^2)}}
    \left(
    \pm i \hat{d}_{j}^+
    \mp
    i \hat{d}_{j}^-
    \right.
    \nonumber\\
    &\quad
    \left.    
    +
    \gamma_{\pm}
    \hat{d}_{j+1}^+
    +
    \gamma_{\pm} 
    \hat{d}_{j+1}^-
    \right)
    ,
\end{align}
where
\begin{align}
    \gamma_{\pm}
    &=
    \frac{1}{4}
    \left[
    \mp
    \frac{t'-t}{\sqrt{tt'}}
    +
    \sqrt{
    16+
    \frac{\left(t-t'\right)^2}{tt'}
    }
    \right]
    .
\end{align}
Let us set $N_c=M+1/2$, i.e., half-integer, so that all the sites in the effective chain $\hat{H}_{l=0,1}^{\text{eff}}$ are coupled through the inter-leg hoppings. By using the CLS operators in Eq.~\eqref{eq:op_W_dd} and also the operators in Eqs.~\eqref{eq:op_LR} for the edges, the Hamiltonian $\hat{H}_{l=0,1}^{\text{eff}}$ in Eq.~\eqref{eq:H_combo_eff} for $\phi=\pi/2$ is diagonalised as
\begin{align} \label{eq:H_combo_eff_Delta12}
    \hat{H}_{l=0,1}^{\text{eff}}
    &=
    \sum_{j=1}^{2M-1}
    \left[
    \left(
    2\sqrt{tt'}\gamma_{+}
    -(-1)^j
    t_j
    \right)
    \hat{W}_{j,+}''^{\dagger} \hat{W}_{j,+}''
    \right.
    \nonumber\\
    &\quad
    \left.
    +
    \left(
    -2\sqrt{tt'}\gamma_{-}
    -(-1)^j
    t_j
    \right)
    \hat{W}_{j,-}''^{\dagger} \hat{W}''_{j,-}
    \right]
    \nonumber\\
    &\quad
    -t'
    \hat{L}^{\dagger}\hat{L} 
    -t
    \hat{R}^{\dagger}\hat{R}
    ,
\end{align}
with $t_{2j-1}=t'$ and $t_{2j}=t$.
The first two lines lead to four flat bands whose energies are $\pm2\sqrt{tt'}\gamma_{\pm}+t'$ and $\pm2\sqrt{tt'}\gamma_{\pm}-t$. The last two states, $\ket{R}$ and $\ket{L}$, are located within energy gaps between flat bands.  

The chain $\hat{H}_{l=0,1}^{\text{eff}}$ in this case can be decomposed, in the symmetric and antisymmetric basis, as
\begin{align} \label{eq:H_combo_eff_t_tp}
    \hat{H}_{l=0,1}^{\text{eff}}
    &=
    \sum_{j=1}^{M-1}
    \bigg[
    \left(
    i\kappa_{1}
    \hat{d}_{2j-1}^{a\dagger} 
    \hat{d}_{2j}^{s}
    +
    \text{H.c.}
    \right)
    \nonumber\\
    &
    +
    \epsilon_{\text{odd}}^a
    \hat{d}_{2j-1}^{a\dagger}
    \hat{d}_{2j-1}^{a}
    +
    \epsilon_{\text{even}}^s
    \hat{d}_{2j}^{s\dagger}
    \hat{d}_{2j}^{s}
    \bigg]
    -
    t\hat{d}_{2M-1}^{a\dagger}\hat{d}_{2M-1}^{a}
    \nonumber\\
    &
    +
    \sum_{j=1}^{M-2}
    \bigg[
    \left(
    i\kappa_{2}
    \hat{d}_{2j}^{a\dagger} 
    \hat{d}_{2j+1}^{s}
    +
    \text{H.c.}
    \right)
    \nonumber\\
    &
    +
    \epsilon_{\text{even}}^a
    \hat{d}_{2j}^{a\dagger}
    \hat{d}_{2j}^{a}
    +
    \epsilon_{\text{odd}}^s
    \hat{d}_{2j+1}^{s\dagger}
    \hat{d}_{2j+1}^{s}
    \bigg]
    -t'\hat{d}_1^{s\dagger}\hat{d}_1^{s}
    ,
\end{align}
where
\begin{subequations}
\begin{align}
    \epsilon_{\text{odd}}^a
    &\equiv
    \frac{t'+2\gamma_{+}\sqrt{tt'}}{1+\gamma_{+}^2}
    +
    \frac{t'-2\gamma_{-}\sqrt{tt'}}{1+\gamma_{-}^2}
    ,
    \\
    \epsilon_{\text{even}}^s
    &\equiv
    \frac{\gamma_{+}^2\left(t'+2\gamma_{+}\sqrt{tt'}\right)}{1+\gamma_{+}^2}
    +
    \frac{-\gamma_{-}^2\left(-t'+2\gamma_{-}\sqrt{tt'}\right)}{1+\gamma_{-}^2}
    ,
    \\
    \kappa_{1}
    &\equiv
    \frac{-\gamma_{+}\left(t'+2\gamma_{+}\sqrt{tt'}\right)}{1+\gamma_{+}^2}
    +
    \frac{-\gamma_{-}\left(-t'+2\gamma_{-}\sqrt{tt'}\right)}{1+\gamma_{-}^2}
\end{align}
\end{subequations}
and
\begin{subequations}
\begin{align}
    \epsilon_{\text{even}}^a
    &\equiv
    \frac{-t+2\gamma_{+}\sqrt{tt'}}{1+\gamma_{+}^2}
    +
    \frac{-t-2\gamma_{-}\sqrt{tt'}}{1+\gamma_{-}^2}
    ,
    \\
    \epsilon_{\text{odd}}^s
    &\equiv
    \frac{\gamma_{+}^2\left(-t+2\gamma_{+}\sqrt{tt'}\right)}{1+\gamma_{+}^2}
    +
    \frac{-\gamma_{-}^2\left(t+2\gamma_{-}\sqrt{tt'}\right)}{1+\gamma_{-}^2}
    ,
    \\
    \kappa_{2}
    &\equiv
    \frac{-\gamma_{+}\left(-t+2\gamma_{+}\sqrt{tt'}\right)}{1+\gamma_{+}^2}
    +
    \frac{-\gamma_{-}\left(t+2\gamma_{-}\sqrt{tt'}\right)}{1+\gamma_{-}^2}
    .
\end{align}
\end{subequations}
The imbalance $t\neq t'$ induces non-uniform on-site energies and breaks the inversion and chiral symmetries. Therefore, the usual 1D topological invariants are either not quantized (Zak phases of the bands) or undefined (winding number) pointing to a lack of topological protection for the persisting in-gap edge states. On the other hand, the presence of four flat bands in the spectrum is seen to be robust against a variation of the $t/t'$ ratio.

\begin{figure}[tb]
    \includegraphics[width=0.95\linewidth]{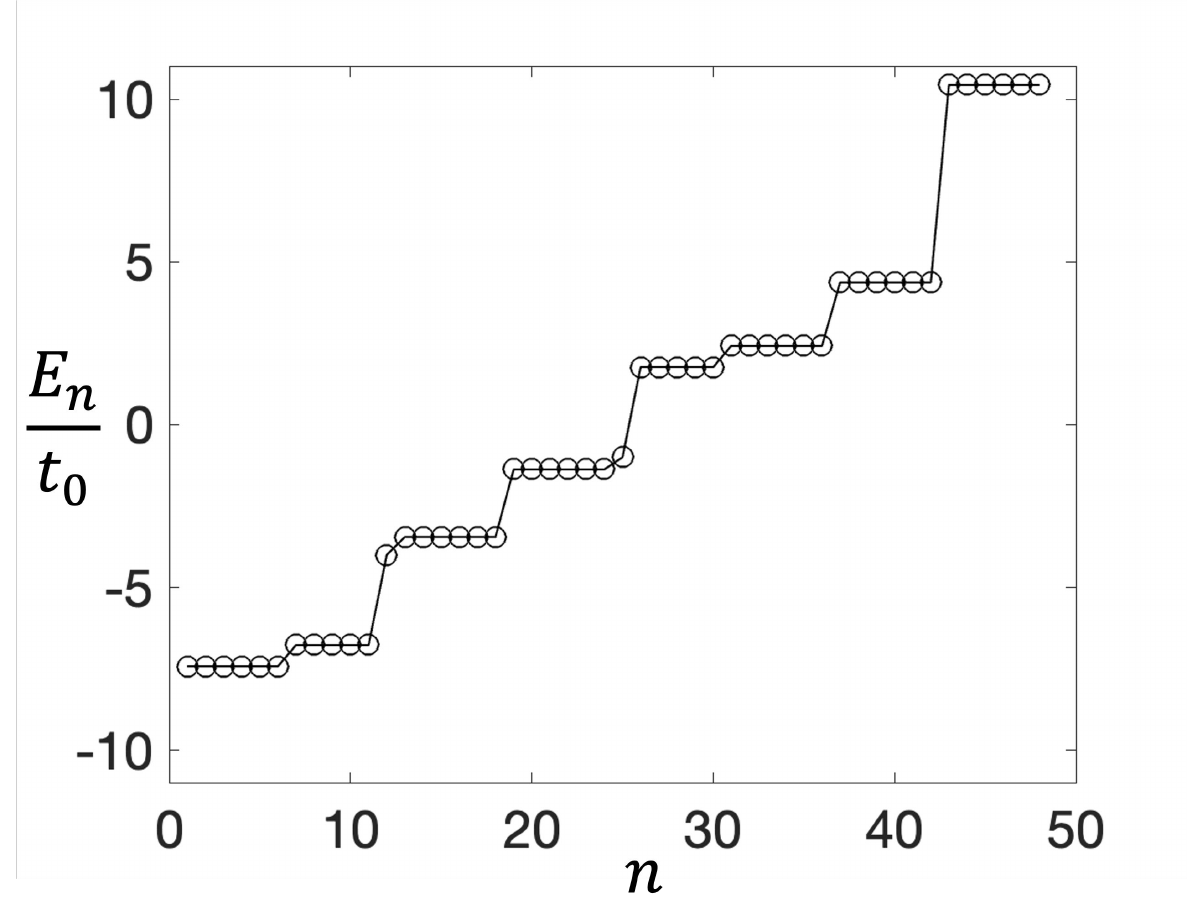}
    \caption{
    Energy spectrum of the effective doublon Hamiltonian in Eq.~\eqref{eq:H_combo_eff_2} for $\{t_j/t_0\}_{j=1}^{2M}=\{1,2,3,4,1,\ldots\}$ and $N_{c}=12.5$ ($M=12$). 
    }
    \label{fig:spectrum_multiplebands}
\end{figure}

\subsubsection{Generation of multiple flat bands} \label{sec:multipleflatbands}

The two vertical hoppings in the previous subsection  alternate at every site, i.e. $\{-t',t,-t',t,\ldots\}$. Following Ref.~\cite{Kuno2020Extended}, we will show that, if the vertical hopping strengths vary with a longer periodicity, such as $\{-t_1,t_2,-t_3,t_4,-t_1,\ldots\}$, more flat bands are generated. 
This can be designed by inducing different interaction strengths and hopping strengths in different cells, for instance by adjusting the distances $d$ and $d'$ between rings. This gives more controllablity to the parameters in the effective model.

We consider the same model as in Eq.~\eqref{eq:H_combo_eff} for $\phi=\pi/2$ but let the vertical hopping strength have an arbitrarily large periodicity. The Hamiltonian now reads
\begin{align} \label{eq:H_combo_eff_tj}
    \hat{H}_{l=0,1}^{\text{eff}}
    &= 
    i
    \sum_{j=1}^{2M-1} 
    \sqrt{t_jt_{j+1}}
    \left[
    \hat{d}_{j+1}^{+\dagger} \hat{d}_{j}^{+} 
    - 
    \hat{d}_{j+1}^{-\dagger} 
    \hat{d}_{j}^{-} 
    \right.
    \nonumber\\
    &\quad\quad\quad\quad\quad\quad\quad\quad
    \left. 
    -
    \hat{d}_{j+1}^{+\dagger} \hat{d}_{j}^{-}
    +
    \hat{d}_{j+1}^{-\dagger} \hat{d}_{j}^{+}
    \right]
    \nonumber\\
    &\quad
    +
    \sum_{j=1}^{2M} 
    (-1)^j t_j
    \hat{d}_{j}^{-\dagger} 
    \hat{d}_{j}^{+}
    +
    \text{H.c.}
    ,
\end{align}
where we fix $N_c=M+1/2$, with $M\in\mathbb{N}$.
We rewrite the CLS operators given in Eq.~\eqref{eq:op_W_dd} such that the coefficients are position dependent,
\begin{align} \label{eq:op_W3}
    \hat{W}'''_{j,\pm}
    &\equiv
    \frac{1}{\sqrt{2(1+\gamma_{j,\pm}^2)}}
    \left(
    \pm i \hat{d}_{j}^+
    \mp
    i \hat{d}_{j}^-
    \right.
    \nonumber\\
    &\quad
    \left.    
    +
    \gamma_{j,\pm}
    \hat{d}_{j+1}^+
    +
    \gamma_{j,\pm} 
    \hat{d}_{j+1}^-
    \right)
    ,
\end{align}
where
\begin{align}
    \gamma_{j,\pm}
    &=
    \frac{1}{4}
    \left[
    \pm(-1)^j
    \left(
    \sqrt{\frac{t_{j}}{t_{j+1}}}
    -
    \sqrt{\frac{t_{j+1}}{t_j}}
    \right)
    \right.
    \nonumber\\
    &\quad
    \left.
    +
    \sqrt{
    16
    +
    \left(
    \sqrt{\frac{t_j}{t_{j+1}}}
    -
    \sqrt{\frac{t_{j+1}}{t_{j}}}
    \right)^2
    }
    \right]
    .
\end{align}
The CLS operators in Eq.~\eqref{eq:op_W3} diagonalise $\hat{H}_{l=0,1}^{\text{eff}}$ in Eq.~\eqref{eq:H_combo_eff_tj} as 
\begin{align}
    \hat{H}_{l=0,1}^{\text{eff}}
    &=
    \sum_{j=1}^{2M-1}
    \left[
    \left(
    2\sqrt{t_jt_{j+1}}\gamma_{j,+}
    -(-1)^j
    t_j
    \right)
    \hat{W}_{j,+}'''^{\dagger} \hat{W}_{j,+}'''
    \right.
    \nonumber\\
    &\quad
    \left.
    +
    \left(
    -2\sqrt{t_jt_{j+1}}\gamma_{j,-}
    -(-1)^j
    t_j
    \right)
    \hat{W}_{j,-}'''^{\dagger} \hat{W}'''_{j,-}
    \right]
    \nonumber\\
    &\quad
    -t_1
    \hat{L}^{\dagger}\hat{L} 
    -t_{2M}
    \hat{R}^{\dagger}\hat{R}
    ,
\end{align}
which contains flat bands with energies $\pm2\sqrt{t_jt_{j+1}}\gamma_{j,\pm}-(-1)^jt_j$ for $j\in[1,2M-1]$ and two non-topological edge states, $\ket{L},\ket{R}$. For instance, the energy spectrum for $\{t_j/t_0\}_{j=1}^{2M}=\{1,2,3,4,1,\ldots\}$, corresponding to a periodicity of four for the hopping terms, and for $M=12$ is plotted in Fig.~\ref{fig:spectrum_multiplebands}. 
The number of flat bands (eight) is the double of the periodicity of the hopping terms (four) because the flat bands are the eigenvalues characterised by $\hat{W}_{j,+}'''$ and $\hat{W}_{j,-}'''$.

\section{Conclusions}
\label{sec:conclusions}

We have investigated doublons arising from NN interactions in a 1D staggered lattice of rings in the hard-core boson limit with: (i) $l=1$ in all the sites and (ii) $l=0$ and $l=1$ in alternating sites. For the case of $l=1$, the physical system can be effectively mapped onto a Creutz ladder using circulations as a synthetic dimension while the effective model for doublon consists of two Bose-Hubbard chains and two SSH chains, opening the possibility for topologically protected edge states. The configuration with $l=0$ and $l=1$ in alternating sites yields an effective model for doublons corresponding to a Creutz ladder with extra inter-leg couplings. 
To the best of our knowledge, a proposal for the implementation of this extended Creutz ladder~\cite{Kuno2020Extended} in a concrete physical platform has been lacking thus far.
Our work provides a way for the realization of a four flat-band extended Creutz ladder for doublon, with staggered vertical tunneling and complex diagonal and horizontal tunnelings. We have examined the topology and the band structure and found that the effective model decouples into two independent SSH chains.
Furthermore, we have investigated cases of different terminations of the original chain and coupling strengths and identified a pathway to generate multiple flat bands in the system. Owing to the presence of four distinct $\pi$ flux loops, our work offers a good platform to study Aharonov–Bohm (AB) caging in a multiple flat-band regime~\cite{PhysRevLett.81.5888}. We have benchmarked the energy spectra of all the effective models against exact numerical solutions and found excellent agreement. 

These findings highlight the rich interplay between lattice geometry, synthetic dimensions, and interaction-induced phenomena, offering pathways for engineering topologically nontrivial phases and flat-band structures in cold atom systems.
Finally, our results are not limited to atomic systems. The realization of synthetic magnetic fields for photons using linear ring resonators  like in Refs.~\cite{Hafezi2013Imaging,Mittal2014,Mittal2016}, together with recent progress on the generation of strong photon–photon interactions~\cite{Roushan2017Chiral,Morvan2022Formation}
, opens possibilities for exploring NN-induced doublons in photonic systems.

\section{Acknowledgments}

We acknowledge Eloi Nicolau for sharing his code and David Viedma for his suggestions.
Y.Z., A.U., and V.A. acknowledge financial support support from the Spanish Ministerio de Ciencia e Innovación (MCIN) (MCIN/AEI/10.13039/501100011033, contract No. PID2020-118153GB-I00) and Generalitat de Catalunya (Contract No. SGR2021-00138). Additionally, Y.Z. and V.A. acknowledge also financial support from Spanish MCIN (MCIN/AEI/10.13039/501100011033, contract No. PID2024-160393NB-I00). Y.Z. acknowledges the China Scholarship Council under the Grant No. 202306890031. A.U. is financially supported by JSPS Overseas Research Fellowships and acknowledges financial support from Spanish MCIN (MCIN/AEI/10.13039/501100011033, contract No. PID2022-141283NB-I00 and No. PID2022-139099NB-I00) with the support of FEDER funds. 
A.M.M. and R.G.D. developed their work within the scope of Portuguese Institute for Nanostructures, Nanomodelling and Nanofabrication (i3N) Projects No. UIDB/50025/2020, No. UIDP/50025/2020, and No. LA/P/0037/2020, financed by national funds through the Fundação para a Ciência e Tecnologia (FCT) and the Ministério da Educação e Ciência (MEC) of Portugal. A.M.M. acknowledges financial support from i3N through the work Contract No. CDL-CTTRI-91-SGRH/2024.

\section*{Data availability}
The data that support the findings of this article are openly available~\cite{Usui2025Code}; embargo periods may apply.
\appendix

\section{Derivation of the effective Hamiltonians for the doublons} \label{app:H_eff_l1}

We expand the Hamiltonian in the manifold of doublon states defined in Eqs.~\eqref{eq:d_ss}-\eqref{eq:d_as} and add the hopping terms as perturbations. Defining the doublon state as $\ket{d_j^{\alpha\beta}}\equiv \hat{d}_j^{\alpha\beta\dagger}\ket{0}$ for $\alpha,\beta=\{s,a\}$, with $\ket{0}$ the vacuum state, and taking into account corrections up to the second-order, each element of the effective Hamiltonian is given by
\begin{align}
    &\mel{d_j^{\alpha\beta}}{\hat{H}_{\text{eff}}}{d_n^{\alpha\beta}}
    \nonumber\\
    &=
    V\delta_{j,n}
    +
    \mel{d_j^{\alpha\beta}}{\hat{H}_{l=1}^0}{d_n^{\alpha\beta}}
    \nonumber\\
    &\quad
    +
    \frac{1}{2}
    \sum_{\alpha\in C}
    \mel{d_j}{\hat{H}_0}{\alpha}
    \mel{\alpha}{\hat{H}_0}{d_{n}}
    \left(
    \frac{1}{V-\epsilon_\alpha} + \frac{1}{V-\epsilon_\alpha}
    \right)
    \nonumber\\
    &=
    V\delta_{j,n}
    +
    \mel{d_j^{\alpha\beta}}{\hat{H}_{l=1}^0}{d_n^{\alpha\beta}}
    \nonumber\\
    &\quad
    +
    \sum_{\alpha\in C}
    \mel{d_j}{\hat{H}_0}{\alpha}
    \mel{\alpha}{\hat{H}_0}{d_{n}}
    \frac{1}{V}
    ,
\end{align}
where $C$ is the set of mediating states where the particles are two sites apart and which have a vanishing interacting energy, $\epsilon_{\alpha}=0$. Also, we define the effective on-site energy and the effective hopping term as $\epsilon_j^{\alpha\beta}\equiv\mel{d_j^{\alpha\beta}}{\hat{H}_{\text{eff}}}{d_j^{\alpha\beta}}$ and $J_{j,n}^{\alpha\beta}\equiv\mel{d_j^{\alpha\beta}}{\hat{H}_{\text{eff}}}{d_n^{\alpha\beta}}$ with $j\neq n$, respectively, as
\begin{align}
    \hat{H}_{\text{eff}}
    &=
    \sum_{j,n=1}^{2N_c-1}
    \ketbra{d_j^{\alpha\beta}}
    \hat{H}_{\text{eff}}
    \ketbra{d_n^{\alpha\beta}}
    \nonumber\\
    &=
    \sum_{j=1}^{2N_c-1}
    \ketbra{d_j^{\alpha\beta}}
    \epsilon_j^{\alpha\beta}
    +
    \sum_{j\neq n}^{2N_c-1}
    \ketbra{d_j^{\alpha\beta}}
    J_{j,n}^{\alpha\beta}
    .
\end{align}

By calculating $\epsilon_j^{\alpha\beta}$ and $J_{j,n}^{\alpha\beta}$, the effective Hamiltonians defined in Eqs.~\eqref{eq:H_eff_l1}-\eqref{eq:epsilon_sa_as} in the main text are derived. 
The effective Hamiltonian in Eq.~\eqref{eq:H_combo_eff} for the case of $l=0$ and $l=1$ alternating sites is also derivated in the same way.
As an example, we show the derivation of $\hat{H}_{\text{eff}}^{as}$ in Eqs.~\eqref{eq:H_sa_as} and~\eqref{eq:epsilon_sa_as}.
Firstly, the on-site energy at odd indices in the bulk is given by
\begin{align}
    \epsilon_{2j-1}^{as}
    &=
    V
    +
    \sum_{\alpha\in C}
    \mel{d_{2j-1}^{as}}{\hat{H}_0}{\alpha}
    \mel{\alpha}{\hat{H}_0}{d_{2j-1}^{as}}
    \frac{1}{V}
    \nonumber\\
    &=
    V
    +
    \left|\mel{b_{j-1}^a b_j^s}{\hat{H}_0}{d_{2j-1}^{as}}\right|^2
    \frac{1}{V}
    \nonumber\\
    &\quad
    +
    \left|\mel{a_{j}^a a_{j+1}^s}{\hat{H}_0}{d_{2j-1}^{as}}\right|^2
    \frac{1}{V}
    \nonumber\\
    &=
    V + \frac{t_a'^2+t_s'^2}{V}
\end{align}
for $3\leq 2j-1 < 2N_c-1$, where site $2N_c-1$ is the right end.
Note that the infinite on-site interaction $U\to\infty$ prohibits double occupation for all sites.
For $j=1$, the process $\mel{b_{j-1}^a b_j^s}{\hat{H}_0}{d_{2j-1}}$ does not exist, and thus the on-site energy at the left edge is $\epsilon_1^{as}=V+t_s'^2/V$.
Secondly, the on-site energy at even indices in the bulk is given by
\begin{align}
    \epsilon_{2j}^{as}
    &=
    V
    +
    \sum_{\alpha\in C}
    \mel{d_{2j}^{as}}{\hat{H}_0}{\alpha}
    \mel{\alpha}{\hat{H}_0}{d_{2j}^{as}}
    \frac{1}{V}
    \nonumber\\
    &=
    V
    +
    \left|\mel{a_{j}^a a_{j+1}^s}{\hat{H}_0}{d_{2j}^{as}}\right|^2
    \frac{1}{V}
    \nonumber\\
    &\quad
    +
    \left|\mel{b_{j}^a b_{j+1}^s}{\hat{H}_0}{d_{2j}^{as}}\right|^2
    \frac{1}{V}
    \nonumber\\
    &=
    V + \frac{t_a^2+t_s^2}{V}
\end{align}
for $2\leq 2j < 2N_c-1$.
The on-site energy at the right edge depends on whether $2N_c-1$ is odd or even. If $2N_c-1$ is odd, we have $\epsilon_{2N_c-1}^{as}=V+t_a'^2/V$. If $2N_c-1$ is even, then $\epsilon_{2N_c-1}^{as}=V+t_a^2/V$.

\begin{figure}[H]
    \includegraphics[width=0.99\linewidth]{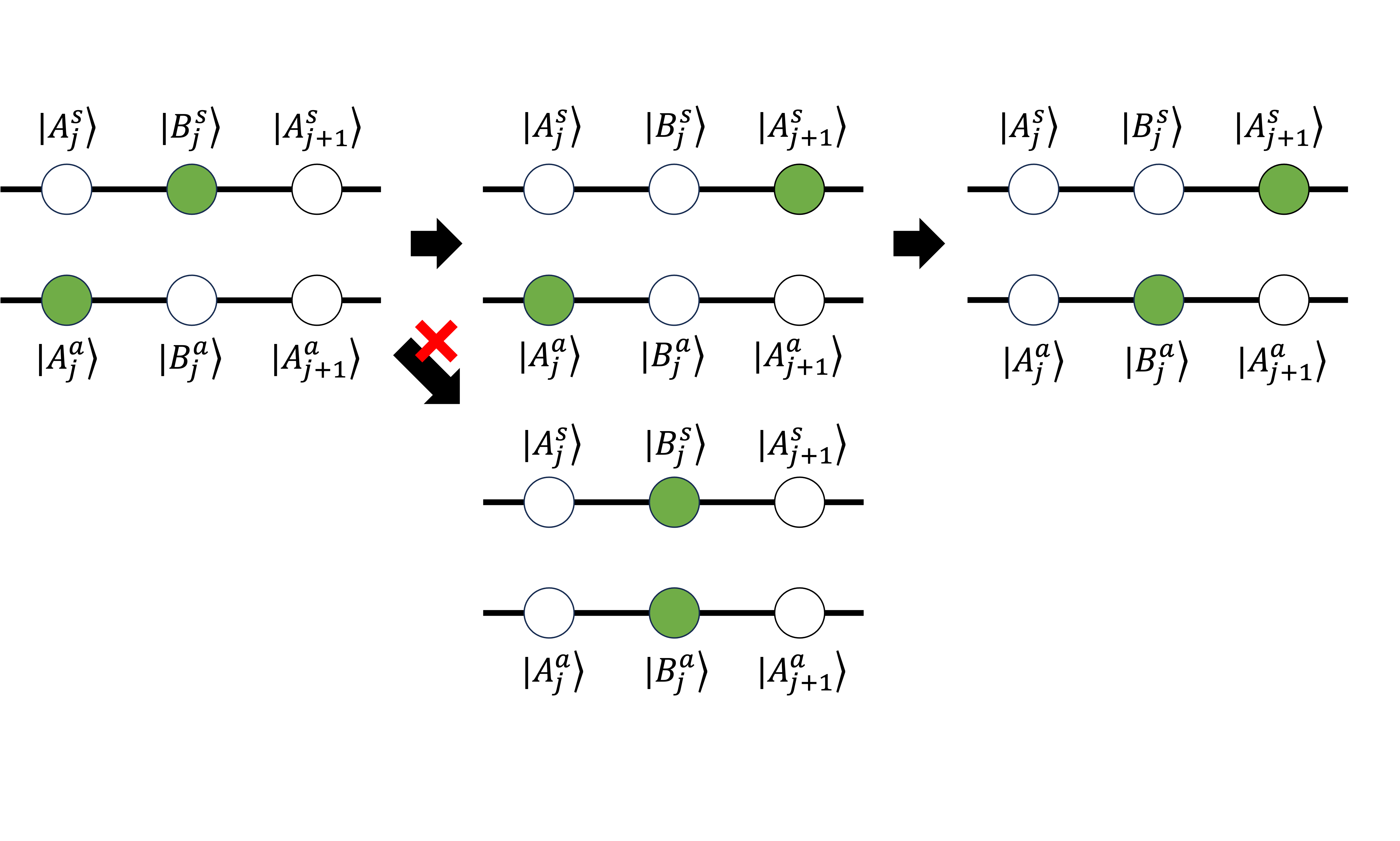}
    \caption{
    Sketch of the second-order processes that determine $\mel{d_{2j}^{as}}{\hat{H}_{\text{eff}}}{d_{2j-1}^{as}}$. 
    The state of the bottom cannot mediate the transition, as it corresponds to a double occupancy that is forbidden by the hard-core constraint $(U\to\infty)$.
    }
    \label{fig:2nd}
\end{figure}

Next, we calculate the effective hopping terms as follows,
\begin{align}
    J_{2j,2j-1}^{as}
    &=
    \mel{d_{2j}^{as}}{\hat{H}_{\text{eff}}}{d_{2j-1}^{as}}
    \nonumber\\
    &=
    \mel{d_{2j}^{as}}{\hat{H}_0}{a_j^a a_{j+1}^s}
    \mel{a_j^a a_{j+1}^s}{\hat{H}_0}{d_{2j-1}^{as}}
    \frac{1}{V}
    \nonumber\\
    &=
    \frac{t_at_s'}{V}
    ,
\end{align}
and
\begin{align}
    J_{2j+1,2j}^{as}
    &=
    \mel{d_{2j+1}^{as}}{\hat{H}_{\text{eff}}}{d_{2j}^{as}}
    \nonumber\\
    &=
    \mel{d_{2j+1}^{as}}{\hat{H}_0}{b_j^a b_{j+1}^s}
    \mel{b_j^a b_{j+1}^s}{\hat{H}_0}{d_{2j}^{as}}
    \frac{1}{V}
    \nonumber\\
    &=
    \frac{t_st_a'}{V}
    .
\end{align}
Therefore, $\hat{H}_{\text{eff}}^{as}$ in Eq.~\eqref{eq:H_sa_as} in the main text is obtained. The second-order process corresponding to the $\mel{d_{2j}^{as}}{\hat{H}_{\text{eff}}}{d_{2j-1}^{as}}$ transition is depicted in Fig.~\ref{fig:2nd}.

\section{\texorpdfstring{Additional studies in case of alternating $l=0$ and $l=1$}{Additional studies in case of alternating l=0 and l=1}}
\label{app:bsite_ending}

In contrast to Sec.~\ref{sec:doublon_l01}, we consider a different case
\begin{figure}[htb]
    \centering
    \includegraphics[width=0.93\linewidth]{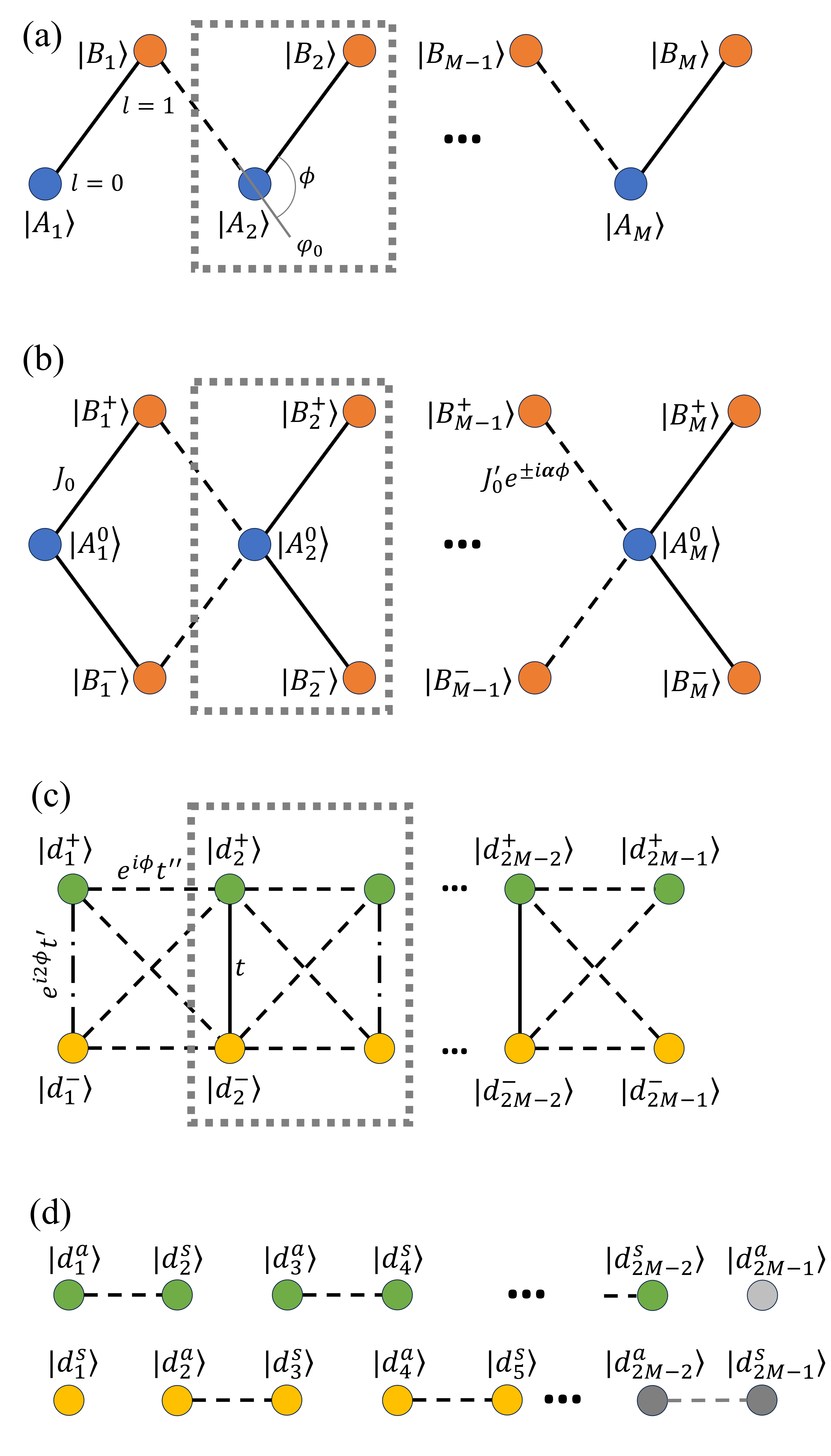}
    \caption{
    (a) Sketch of the 1D staggered chain of rings loaded with atoms with OAM $l=0$ and $l=1$, ending up with a $B$ site such that the number of unit cells is integer, $N_c=M$ with $M\in\mathbb{N}$.
    (b) Sketch of a diamond chain in the real dimension and the synthetic dimension spanned by the circulations $\pm$.
    (c) Sketch of the effective model~\eqref{eq:H_combo_eff} for the doublon subspace with annihilation operators given by Eq.~\eqref{eq:doublons_diamond}.
    The solid lines indicate real tunneling amplitudes, and the dashed and dashed-dotted lines indicate complex tunneling amplitudes.
    The number of sites here is $2N_c-1=2M-1$. In contrast to Fig.~\ref{fig:l01model}(c), the vertical rung coupling in the effective model at the right edge is missing.
    (d) Decomposition of the effective doublon model in Eq.~\eqref{eq:H_combo_eff_1_diag} into two decoupled SSH chains.
    The grey-coloured line and circles on the right indicate that its coupling and their on-site energies are different from the rest of the system. 
    }
    \label{fig:diamond_broken}
\end{figure}
where the original chain ends with a $B$ site, as shown in Figs.~\ref{fig:diamond_broken}(a) and (b), such that $N_c=M$ with $M\in\mathbb{N}$. In this case, the coupling between the doublons $\ket{d_{2M-1}^+}$ and $\ket{d_{2M-1}^-}$ at the right edge is suppressed [see Fig.~\ref{fig:diamond_broken}(c)]. Because of this, we need to modify the CLSs near the right edge, and thus we replace $\hat{W}_{2M-2,\pm}$ by the following,
\begin{align} \label{eq:op_CLS_prime}
    \hat{W}_{2M-2,\pm}'
    &\equiv
    \frac{1}{\sqrt{2(1+\gamma_{R,\pm}^2)}}
    \left(
    \pm i \hat{d}_{2M-2}^{+}
    \mp
    i \hat{d}_{2M-2}^{-}
    \right.
    \nonumber\\
    &\quad\quad\quad
    \left.
    +
    \gamma_{R,\pm}
    \hat{d}_{2M-1}^{+}
    +
    \gamma_{R,\pm} 
    \hat{d}_{2M-1}^{-}
    \right)
    ,
\end{align}
with $\gamma_{R,\pm}=(\sqrt{17}\pm1)/4$,
where the population imbalance is induced between sites $2M-2$ and $2M-1$ by $\gamma_{R,\pm}$.
\par 
By applying the set of CLS operators in Eqs.~\eqref{eq:op_CLS}, \eqref{eq:op_LR} and \eqref{eq:op_CLS_prime}, the effective Hamiltonian $\hat{H}_{l=0,1}^{\text{eff,}1}$ is diagonalised as
\begin{align} \label{eq:H_combo_eff_open}
    \hat{H}_{l=0,1}^{\text{eff,}1}
    &=
    \sum_{j=1}^{2M-3}
    \left[
    t
    \left(
    2-(-1)^j
    \right)
    \hat{W}_{j,+}^{\dagger} \hat{W}_{j,+}
    \right.
    \nonumber\\
    &\quad
    \left.
    +
    t
    \left(
    -2-(-1)^j
    \right)
    \hat{W}_{j,-}^{\dagger} \hat{W}_{j,-}
    \right]
    -
    t
    \hat{L}^{\dagger}\hat{L} 
    \nonumber\\
    &\quad
    +
    t
    \left(
    2\gamma_{R,+}-1
    \right)
    \hat{W}_{2M-2,+}'^{\dagger} \hat{W}_{2M-2,+}'
    \nonumber\\
    &\quad
    +
    t
    \left(
    -2\gamma_{R,-}-1
    \right)
    \hat{W}_{2M-2,-}'^{\dagger} \hat{W}_{2M-2,-}'
    \nonumber\\
    &\quad
    +
    0~
    \hat{R}^{\dagger}\hat{R}
    .
\end{align}
The first two lines are the same as the previous case in Eq.~\eqref{eq:H_combo_eff_2} and generate four flat bands with energies $-3t,-t,t,3t$. The lack of the vertical hopping at the right edge shifts the energy of the three states localised near the right edge, $\ket{W_{2M-2,+}'},\ket{W_{2M-2,-}'},\ket{R}$, from the flat bands to $t(2\gamma_{R,+}-1)$, $t(-2\gamma_{R,-}-1)$, $0$. These states are not topological states but impurity states, as we explain below. 
We rewrite Eq.~\eqref{eq:H_combo_eff_open} in the symmetric and antisymmetric basis given in Eq.~\eqref{eq:op_d_sa},
\begin{align} \label{eq:H_combo_eff_open_CLS}
    \hat{H}_{l=0,1}^{\text{eff,}1}
    &=
    \sum_{j=1}^{M-1}
    \bigg[
    \left(
    -2it
    \hat{d}_{2j-1}^{a\dagger} 
    \hat{d}_{2j}^{s}
    +
    \text{H.c.}
    \right)
    \nonumber\\
    &\quad
    +t
    \left(
    \hat{d}_{2j-1}^{a\dagger}
    \hat{d}_{2j-1}^{a}
    +
    \hat{d}_{2j}^{s\dagger}
    \hat{d}_{2j}^{s}
    \right)
    \bigg]
    +
    0~\hat{d}_{2M-1}^{a\dagger}\hat{d}_{2M-1}^{a}
    \nonumber\\
    &
    +
    \sum_{j=1}^{M-2}
    \bigg[
    \left(
    -2it
    \hat{d}_{2j}^{a\dagger} 
    \hat{d}_{2j+1}^{s}
    +
    \text{H.c.}
    \right)
    \nonumber\\
    &\quad\quad
    -t
    \left(
    \hat{d}_{2j}^{a\dagger}
    \hat{d}_{2j}^{a}
    +
    \hat{d}_{2j+1}^{s\dagger}
    \hat{d}_{2j+1}^{s}
    \right)
    \bigg]
    \nonumber\\
    &
    +
    \epsilon_{2M-2}^a
    \hat{d}_{2M-2}^{a\dagger}
    \hat{d}_{2M-2}^{a}
    +
    \epsilon_{2M-1}^s
    \hat{d}_{2M-1}^{s\dagger}
    \hat{d}_{2M-1}^{s}
    \nonumber\\
    &
    +
    i\kappa_{2M-2}
    \hat{d}_{2M-2}^{a\dagger}
    \hat{d}_{2M-1}^{s}
    -t\hat{d}_1^{s\dagger}\hat{d}_1^{s}
    ,
\end{align}
where $\epsilon_{2M-2}^a,\epsilon_{2M-1}^s$, and $\kappa_{2M-2}$ are extra on-site energies and hopping term near the right edge, respectively, and are given by
\begin{subequations}
\begin{align}
    \epsilon_{2M-2}^a
    &\equiv
    \left(
    \frac{-1+2\gamma_{R,+}}{1+\gamma_{R,+}^2}
    +
    \frac{-1-2\gamma_{R,-}}{1+\gamma_{R,-}^2}
    \right)
    t,
    \\
    \epsilon_{2M-1}^s
    &\equiv
    \left(
    \frac{\gamma_{R,+}^2(-1+2\gamma_{R,+})}{1+\gamma_{R,+}^2}
    +
    \frac{-\gamma_{R.-}^2(1+2\gamma_{R,-})}{1+\gamma_{R,-}^2}
    \right)
    t,
    \\
    \kappa_{2M-2}
    &\equiv
    \left(
    \frac{-\gamma_{R,+}(-1+2\gamma_{R,+})}{1+\gamma_{R,+}^2}
    +
    \frac{-\gamma_{R,-}(1+2\gamma_{R,-})}{1+\gamma_{R,-}^2}
    \right)
    t
    .
\end{align}
\end{subequations}
The first and the second lines of Eq.~\eqref{eq:H_combo_eff_open_CLS} correspond to an SSH chain with an impurity at the right edge, and the rest of the lines correspond to another SSH chain with extra on-site energies $\epsilon_{2M-2}^a,\epsilon_{2M-1}^s$ and coupling $\kappa_{2M-2}$ near the right edge. Therefore, the three in-gap states $\ket{W_{2M-2,+}'},\ket{W_{2M-2,-}'},\ket{R}$ are originated from the on-site energy imbalance in the SSH chains. Nevertheless, the left edge is not affected, and thus the left edge state $\ket{L}$ remains a topological state.

\bibliography{bibliografia}

\end{document}